%% file: ms.tex
\documentstyle[prl,aps,epsf,epsfig]{revtex} 
\begin{document}
\draft
\twocolumn[\hsize\textwidth\columnwidth\hsize\csname@twocolumnfalse\endcsname
%
\title{Gravitational dependence of contact angle hysteresis \\ for 
a solid defect pinning a liquid-vapor interface}
%
\author{
J\o rgen Vitting Andersen and Henrik Jeldtoft Jensen, }
\address{
Department of Mathematics, Imperial College, Huxley Building, \\ 
180 Queen's Gate, London SW7 2BZ, United Kingdom}
\maketitle
\centerline{\small (Last revised \today)}

\begin{abstract}

We study the case of contact angle hysteresis, when a solid defect 
is quasi-statically pulled out and pushed into a liquid,  
in the presence of gravity.  We solve the problem analytically in an approximation
which, in the case of a truncated parabolic pinning force,
 becomes exact in the the limit of  slow variations in the height of the liquid
vapour interface. The case of a Gaussian pinning force is solved by a 
geometrical construction. We show that the extension of the defect, and not 
the capillary length, is the determining length scale, when the
extension of the defect is much smaller than the capillary length.

\end{abstract}
\vskip2pc]
   
The spreading of a liquid on a solid is 
known to be strongly influenced  by chemical inhomogeneities and  
roughness of the solid surface\cite{deGennes}. 
Different situations arise for the motion of the co--existence line between
a solid, a liquid and a gas phase 
(called the triple line or the contact line)
depending on whether the liquid completely wets the
solid surface \cite{Adelman,completewetting} or
incompletely wets the solid [3-18]. In the case of incomplete 
wetting, which we will consider in this letter, the onset of hysteresis 
for single defect {\em without} gravity was 
first solved analytically in \cite{Joanny,Raphael}. Much effort, 
experimentally [3-10] as well as theoretically [11-19]. 
has since then been done to understand the case with gravity as 
well as the case where many defects pin the triple line. 

Here we study contact 
angle hysteresis for a single Gaussian defect in the presence of gravity. 
Expanding the Gaussian pinning force to second order, we
derive an analytic expression for the  contact angle hysteresis.
In particular we find  a weak {\it logarithmic} dependence on gravity   
when the extension of the defect is small compared to the 
capillary length. In this limit the hysteresis is mainly
determined by the ratio between the liquid-vapor surface energy,  
and the strength of the pinning potential times the extension of the defect. 
Generalising
to the case where the defect is placed on a solid plate, we show that 
the hysteresis then has an additional dependence of the  
capillary length because of the capillary rise along the solid plate. 

We use a  macroscopic Hamiltonian for the 
system consisting of a (horizontal) liquid and a solid vertical plate with a 
fiber of defects.\cite{Joergen}  Let the liquid--vapor interface,
 $h(x,y)$, be chosen along the
$x$-$y$--direction, with the solid being pulled vertically out of the liquid in 
the $z$--direction, see Fig. 1a. Minimizing the Hamiltonian leads to the force balance: 
\begin{eqnarray}
\rho g h(x,y) - \gamma_{LG} \nabla^2 h(x,y) 
(1 + (\vec{\nabla} h(x,y))^2)^{-{3 \over 2}} & =& \nonumber  \\ 
A_p^{'} \Delta y \delta (x) \delta (y) + A_p^{0} \delta (x) & &
\label{forcebalance}
\end{eqnarray}
$\gamma_{LG}$ is the liquid--gas surface energy, $A_p^{'}$ the strength of the 
pinning potential of the
fiber defect and $A_p^0$ is the strength of the pinning potential 
of the plate \cite{note1}. $\rho$ is the mass density of the liquid, $g$ the gravity 
constant and $\Delta y$ the width of the fiber of defects. 
Solving Eq.~(\ref{forcebalance}) in the limit $(\vec{\nabla} h)^2 
\ll 1$ gives:  
\begin{eqnarray}
h(x,y) & =& {A_p^{'} \Delta y \over 2 \pi \gamma_{LG}} 
 K_0({\sqrt{2} \sqrt{x^2+y^2} \over l_c} ) \nonumber \\
& & + {A^0_p l_c \over 2 \sqrt{2}  \gamma_{LG} } 
\exp{(- {\sqrt{2} x \over l_c} )}
\label{h}
\end{eqnarray}
with $l_c= \sqrt{{2 \gamma_{LG} \over \rho g}}$ the capillary length 
and $K_0$ the zeroth-order modified Bessel function. 

We consider first the case without plate-liquid interaction, $A_p^0 \equiv 0$. 
In this case the problem is rotational symmetric around the axis of the fiber. 
The force per unit length (for a given width, $\Delta y$)
needed to lift the liquid the distance $h(0,\Delta y)$ along the fiber is 
determined by Eq.~(\ref{h}) as,
\begin{eqnarray}
A_p^{'}& =& {h(0,\Delta y) 2 \pi \gamma_{LG} 
 \over \Delta y K_0({\Delta y \sqrt{2} \over l_c}  )} 
\label{A_p}
\end{eqnarray}
Dispite the fact that  Eq.~(\ref{A_p}) is obtained for a fiber of width 
$\Delta y$, the equation also describes the case of a defect of width $\Delta y$
and zero extension in the $z$-direction. 
This is because the part of the fiber which is above/below the triple line does not 
enter in the force balance of the triple line described by $h(x,y)$. 
The pinning force is in balance with the
gravitational force determined by  the weight of the 
liquid meniscus $F_g = \rho g V$. 
The volume $V$ of the liquid meniscus is given by an integration over 
Eq.~(\ref{h}) with the assumed $A_p^{'}$ from Eq.~(\ref{A_p}). One finds 
\begin{eqnarray}
F_g& =& {2 \pi \gamma_{LG} \over K_0({\Delta y \sqrt{2} \over l_c }) }
h(0,\Delta y) \equiv k h(0,\Delta y)
\label{F_g}
\end{eqnarray}
where we introduced the effective elasticity constant $k$.
Asumming the pinning force is Gaussian, with it's center $z_d$ above $z=0$, 
\begin{equation}
F_p(z) \equiv A_p \Delta y \exp{(-[(z-z_d)/R]^2)},
\label{gauss-pin}
\end{equation}
one can like   
\cite{Joanny} determine the critical amplitude $A_p^c$ for onset of hysteresis
by equating the maximum slope of $F_p(z)$ to the slope of $F_g(z)$:  
\begin{eqnarray}
A_p^{c}& =& {\sqrt{2e} \pi  \gamma_{LG} R 
 \over \Delta y K_0({\sqrt{2} \Delta y \over l_c} )} 
\label{A_c}
\end{eqnarray}
Experimentally, hysteresis is defined as the difference between advancing 
and receding work done on the contact line. We define accordingly the hysteresis 
$H$ by,   
\begin{eqnarray}
H & =& \int_{- \infty}^{+ \infty} F_p(z_t (z_d)) dz_d
+ \int^{- \infty}_{+ \infty} F_p(z_t (z_d)) dz_d
\label{Hysdef}
\end{eqnarray}
where $z_t(z_d)$ describes the dependence of the position of the triple 
line as a function of the position of the defect $z_d$. Notice $H$ can be 
different from $0$ since $F_p$ is not necessarily a single valued function of $z_d$. 
This is due to discontinuities in the function $z_t(z_d)$ when $A_p>A^c_p$.
 As depicted in Fig. 1b, 
$z_t(z_d)$ undergoes a jump from position $z_1$ to $z_2$ for decreasing $z_d$
and a jump from $z_3$ to $z_4$ for increasing $z_d$.
The positions $z_1, z_3$ are determined by the condition $d F_p(z)/dz =k$ and 
the positions $z_2, z_4$ are determined by the condition $k z = F_p (z)$. 
The integrals in Eq.~(\ref{Hysdef}) can be transformed into integrals over
$z_t$ by use of the Jacobiant $dz_t/dz_d=(k-dF_p/dz_t)/k$\cite{note2}
and we can express the hysteresis in terms of the jump positions $z_n$,
$n=1,\cdots,4$. 
\begin{eqnarray}
H & = &  \sum_{n=1}^4 (-1)^{n+1}{1\over 2k}F_p^2(z_n) \nonumber \\ 
& & + \int^{z_2}_{z_3} F_p (z) dz
- \int^{z_1}_{z_4} F_p (z) dz   .
\label{Hys-jump}
\end{eqnarray}
In the case of a Gaussian force Eq.~(\ref{gauss-pin}),   
it is easy numerically 
to find the jump  positions $z_n$ by solving  the geometric construction 
sketched in Fig.~1b. The resulting hysteresis calculated from Eq.~(\ref{Hys-jump}) 
is shown in Fig.~2 versus ${k \over k^*}$, where $k^* = A_p \sqrt{2}/(R \sqrt{e})$ 
is the maximal elasticity constant giving hysteresis. The inset of 
Fig.~2 shows the jump positions $z_n$. Notice that the hysteresis 
diverges as $1/k$ when ${k \over k^*} \rightarrow 0$, because $z_3 \rightarrow 0$ 
while at the same time $|z_1|,|z_2|,|z_4|  \rightarrow \infty$.

A closed analytic form for the hysteresis can be obtained if we replace 
the Gaussian pinning force in Eq.~(\ref{gauss-pin}) by a truncated parabola
\begin{equation}
F_p(z) = \left\{\begin{array}{ll}
 A_p\Delta y[1-({z-z_d\over R})^2]&\mbox{if $|z-z_d|<R$}\\
0                                           &\mbox{otherwise}
                     \end{array}
            \right.
\label{parab-pin}
\end{equation}      
In this case Eq.~(\ref{Hys-jump}) leads to a hysteresis given by 
\begin{eqnarray}
H(\gamma,\rho g,\Delta y,R) 
 = A_p^{'} \Delta y [{1\over 2 C}  
-{3 R^2 C\over 4} 
+ {C^{2} R^{3} \over 2} \nonumber \\ 
- {3 C^3 R^4  \over 32}+{R^4 \over 12}(C^2+4/R^2-4C/R)^{3/2} ] & &
\label{Hys}
\end{eqnarray}
where $C \equiv k/(A_p \Delta y)$. 
The effect of the gravitational field enters through $k$ in $C$. 
Since the zero order Bessel function $K_0(x)$ 
depends logarithmically upon $x$ for small $x$ we conclude that 
when $\Delta y \ll l_c$  the hysteresis  only depends logarithmically
upon the acceleration of gravity, i.e., a very weak dependence. 
In this case the extension of the defect $(\Delta y,R)$ becomes the 
relevant length scale. 
Notice that this is in strong contrast to the common belief that 
the capillary length is the determining length scale in contact 
angle hysteresis. Contrary, when the defect is spatially {\em extended} 
(e.g. in the case of a wall with $A_p^0 \neq 0$),  we show below 
that the hysteresis depends on powers of $l_c$, i.e. a strong 
dependence of gravity.

Like the Gaussian case we have plotted the hysteresis for the 
parabolic force in Fig.~2 versus ${k \over k^*}$. One notices 
that the two different pinning forces give rise to same the $1/k$  
divergence for a weak elastic constant ${k \over k^*} 
\rightarrow 0$, since in this limit the hysteresis is determined 
by the jump solution $z_3$, which coincide for the two different 
shapes of pinning force. On the other hand for ${k \over k^*}
\rightarrow 1$ all jump solutions $z_n$ become relevant in 
determining the hysteresis, and the specific form of the pinning 
force matters. 

In order to illustrate our findings for a typical set of experimental 
values, we have plotted in Fig.~3 the hysteresis versus gravity $\rho g$ 
for different values of defect interaction range $R$. We have used 
a $\gamma_{LG}$ corresponding to water--air, and assumed a contact angle 
hysteresis of the defect of $20^0$. We have furthermore assumed a 
fixed given aspect ratio of the defect given by $R/\Delta y =0.1$. 
The curves $(e)-(h)$ illustrate the very weak dependence of hysteresis 
on gravity when one is not in a parameter regime close to onset of
hysteresis ($\rho g \approx 1000 g/(cm^2s^2)$ corresponds to the density 
of water). For comparison note that a change in $R$ by one decade makes a change
in $H$ by two decades. 
On the other hand when $l_c \approx \Delta y$ a pronounced 
dependence of gravity happens as illustrated by $(a)-(d)$. For a fixed 
$\rho g$ one notices from the curves $(b)$ and $(d)$ (or from Eq.~(\ref{Hys}))
that $H$ goes through a maximum as a function of  
$R$. The same trend happens for the Gaussian pinning force as 
can be seen from the curves $(a)$ and $(c)$.

Let us finally turn our attention to the  general case with $A_p^{0} \neq 0$.
We can repeat the procedure above. 
The gravitational restoring force for system of a plate with a fiber is
given by 
\begin{eqnarray}
F_g& =& {2 \pi \gamma_{LG} \over K_0({\Delta y \sqrt{2} \over l_c }) }
h_0(0,\Delta y) - {A_p^{0} l_c \pi \over \sqrt{2}  
K_0({\Delta y \over \sqrt{2} l_c }) }
+ L_y A_p^{0}
\label{F_ggeneral}
\end{eqnarray}
$L_y$ is the length of the system in the $y$ direction. The last term 
on the right hand side of Eq.~(\ref{F_ggeneral}) is the contribution 
from the solid plate alone, whereas the negative term enters because 
the gravitational force on the fiber is now changed due to the 
additional capillary rise of the plate.  
The hysteresis in the general case with 
a plate plus fiber can be found analytically in the case of a pinning 
force given by Eq.~(\ref{parab-pin}):
\begin{eqnarray}
H(\gamma,\rho g,\Delta y,R) =  & & \nonumber \\
A_p^{'} \Delta y [{1 \over C} ({1\over 2} - C_1)
-C R^2({3 \over 4}+{C_1 \over 4}) - {3 C^3 R^4  \over 32} & & \nonumber \\ 
+{R^4 \over 12}[(C^2+4/R^2+4C_1/R^2-4C/R)^{3/2} 
-({4 C_1 \over R^2})^{3/2}] & &
\label{Hysgeneral}
\end{eqnarray}
where $C_1 \equiv A_p^{0} l_c \pi / [\sqrt{2} K_0({\Delta y \sqrt{2} \over l_c} )
A_p^{'} \Delta y ]$. The linear dependence of $C_1$ on the capillary length $l_c$ 
leads to a leading dependence on gravity in the form $(\rho g)^{-1/2}$.

\underline{\sl Conclusion:}
We have studied the case of contact angle hysteresis, when a solid defect with 
a Gaussian pinning force is quasi-statically pulled out and pushed into a liquid,  
in the presence of gravity.  Expanding the Gaussian pinning force to second order 
and assuming 
slow variations in the height of the liquid
vapour interface, we have derived an analytical expression for the 
contact angle hysteresis. 
In strong contrast to the common belief, we have shown that the 
extension of the defect, and not the  
capillary length, is the determining length scale, when the 
extension of the defect is much smaller than the capillary lenth. 

J.V.A. would like to thank Lyd\'eric Bocquet and Elisabeth Charlaix for 
helpful discussions. 
J.V.A. wishes to acknowledge support from the European Union 
Training and Mobility Programme
under contract number ERBFMBICT960965.

\onecolumn
\widetext
\begin{figure}
\begin{center}
\input{fig1a.tex}
\caption{Fig.~1a}
\end{center}
\end{figure}
\setcounter{figure}{0}
\begin{figure}
\begin{center}
\input{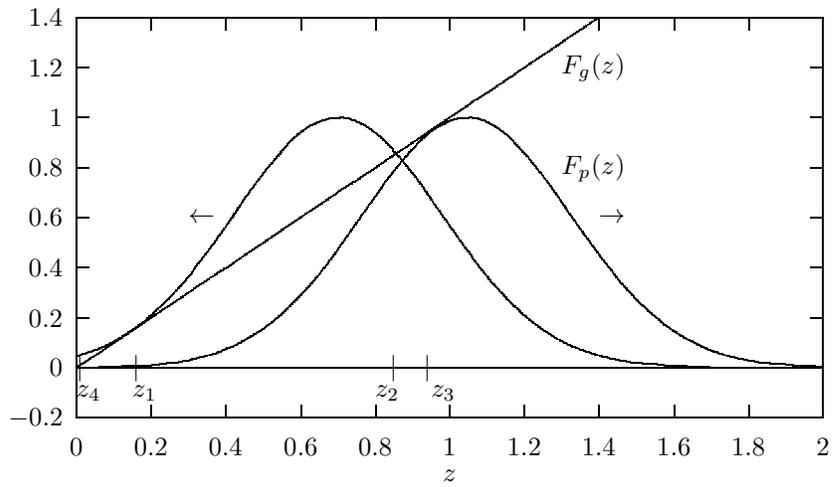}
\caption{$Fig.~1b. F_p$, $F_g$ versus $z$}
\end{center}
\end{figure}
\setcounter{figure}{1}
\begin{figure}
\begin{center}
\input{fig2.tex}
\caption{Hysteresis $H$ versus ${k \over k^*}$ determined for a Gaussian 
defect force using Eq.~(\ref{Hys-jump}) (thick solid line) and for a truncated parabolic 
defect force  using Eq.~(\ref{Hys}) (thin solid line). Inset shows jump positions $z_n$ 
versus ${k \over k^*}$}
\end{center}
\end{figure}
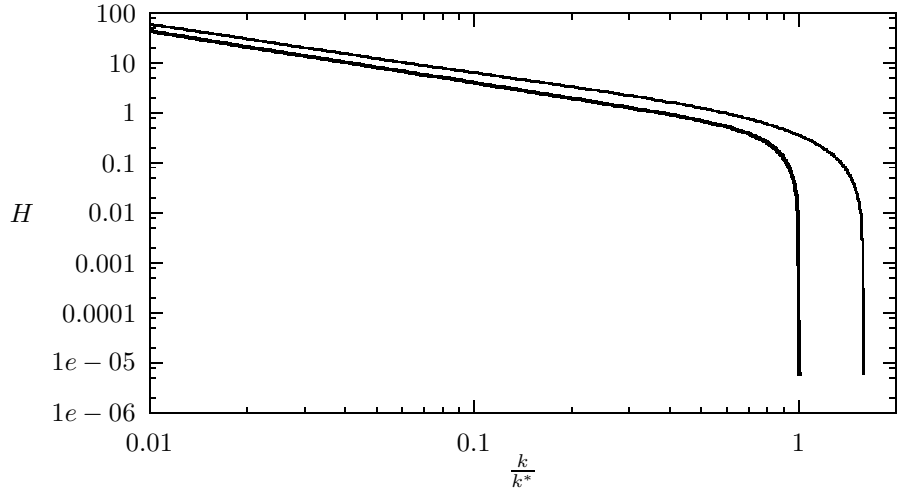
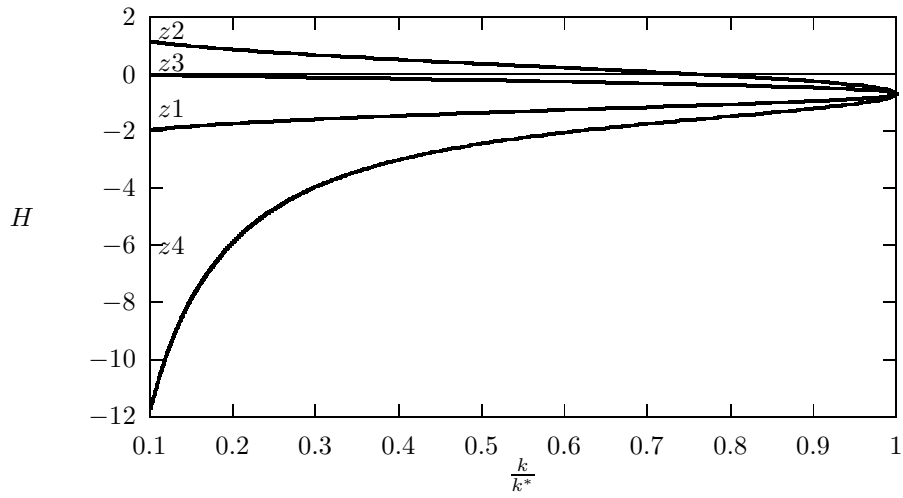
\begin{figure}
\setcounter{figure}{1}
\begin{center}
\input{fig2_inset.tex}
\caption{Inset to Fig.~2}
\end{center}
\end{figure}
\begin{figure}
\begin{center}
\input{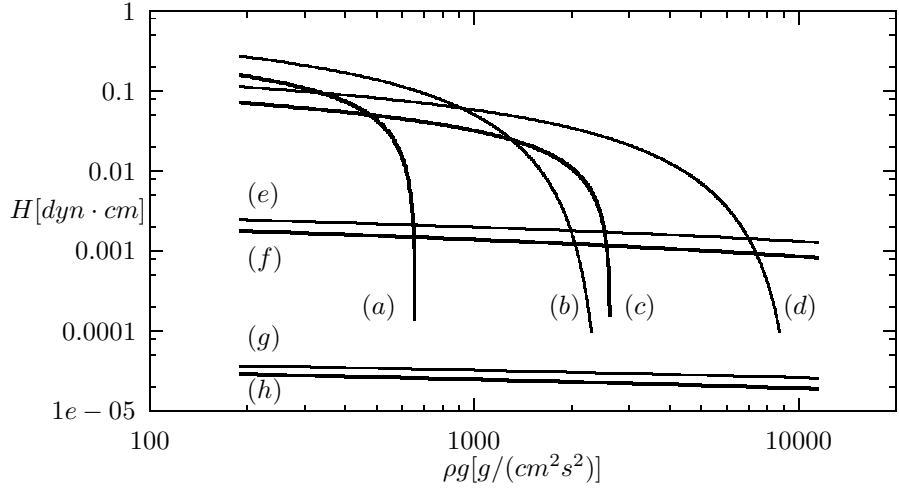}
\caption{Hysteresis $H$ versus $\rho g$ for different values of interaction range $R$ 
determined for a Gaussian defect force using Eq.~(\ref{Hys-jump}) (thick solid lines)
and for a truncated parabolic defect force  using Eq.~(\ref{Hys}) (thin solid lines). 
We have used $\gamma_{LG}=73 dyn/cm$ 
corresponding to water, and a typical experimental value, $20^0$, for contact 
angle hysteresis of the defect, giving $A_p^{'}=69 dyn/cm$. We have used a 
fixed aspect ratio of the defect $R/\Delta y = 0.1$
(a-b) $R=0.02$ cm, (c-d) $R=0.01$ cm, (e-f) $R=0.001$ cm, (g-h) $R=0.0001$ cm.}
\end{center}
\end{figure}
\twocolumn

\end{document}

%% file: fig2.tex
\setlength{\unitlength}{0.240900pt}
\ifx\plotpoint\undefined\newsavebox{\plotpoint}\fi
\sbox{\plotpoint}{\rule[-0.175pt]{0.350pt}{0.350pt}}%
\begin{picture}(1500,900)(0,0)
\sbox{\plotpoint}{\rule[-0.175pt]{0.350pt}{0.350pt}}%
\put(264,158){\rule[-0.175pt]{4.818pt}{0.350pt}}
\put(242,158){\makebox(0,0)[r]{$1e-06$}}
\put(1416,158){\rule[-0.175pt]{4.818pt}{0.350pt}}
\put(264,182){\rule[-0.175pt]{2.409pt}{0.350pt}}
\put(1426,182){\rule[-0.175pt]{2.409pt}{0.350pt}}
\put(264,213){\rule[-0.175pt]{2.409pt}{0.350pt}}
\put(1426,213){\rule[-0.175pt]{2.409pt}{0.350pt}}
\put(264,229){\rule[-0.175pt]{2.409pt}{0.350pt}}
\put(1426,229){\rule[-0.175pt]{2.409pt}{0.350pt}}
\put(264,237){\rule[-0.175pt]{4.818pt}{0.350pt}}
\put(242,237){\makebox(0,0)[r]{$1e-05$}}
\put(1416,237){\rule[-0.175pt]{4.818pt}{0.350pt}}
\put(264,260){\rule[-0.175pt]{2.409pt}{0.350pt}}
\put(1426,260){\rule[-0.175pt]{2.409pt}{0.350pt}}
\put(264,292){\rule[-0.175pt]{2.409pt}{0.350pt}}
\put(1426,292){\rule[-0.175pt]{2.409pt}{0.350pt}}
\put(264,308){\rule[-0.175pt]{2.409pt}{0.350pt}}
\put(1426,308){\rule[-0.175pt]{2.409pt}{0.350pt}}
\put(264,315){\rule[-0.175pt]{4.818pt}{0.350pt}}
\put(242,315){\makebox(0,0)[r]{$0.0001$}}
\put(1416,315){\rule[-0.175pt]{4.818pt}{0.350pt}}
\put(264,339){\rule[-0.175pt]{2.409pt}{0.350pt}}
\put(1426,339){\rule[-0.175pt]{2.409pt}{0.350pt}}
\put(264,370){\rule[-0.175pt]{2.409pt}{0.350pt}}
\put(1426,370){\rule[-0.175pt]{2.409pt}{0.350pt}}
\put(264,386){\rule[-0.175pt]{2.409pt}{0.350pt}}
\put(1426,386){\rule[-0.175pt]{2.409pt}{0.350pt}}
\put(264,394){\rule[-0.175pt]{4.818pt}{0.350pt}}
\put(242,394){\makebox(0,0)[r]{$0.001$}}
\put(1416,394){\rule[-0.175pt]{4.818pt}{0.350pt}}
\put(264,418){\rule[-0.175pt]{2.409pt}{0.350pt}}
\put(1426,418){\rule[-0.175pt]{2.409pt}{0.350pt}}
\put(264,449){\rule[-0.175pt]{2.409pt}{0.350pt}}
\put(1426,449){\rule[-0.175pt]{2.409pt}{0.350pt}}
\put(264,465){\rule[-0.175pt]{2.409pt}{0.350pt}}
\put(1426,465){\rule[-0.175pt]{2.409pt}{0.350pt}}
\put(264,473){\rule[-0.175pt]{4.818pt}{0.350pt}}
\put(242,473){\makebox(0,0)[r]{$0.01$}}
\put(1416,473){\rule[-0.175pt]{4.818pt}{0.350pt}}
\put(264,496){\rule[-0.175pt]{2.409pt}{0.350pt}}
\put(1426,496){\rule[-0.175pt]{2.409pt}{0.350pt}}
\put(264,527){\rule[-0.175pt]{2.409pt}{0.350pt}}
\put(1426,527){\rule[-0.175pt]{2.409pt}{0.350pt}}
\put(264,544){\rule[-0.175pt]{2.409pt}{0.350pt}}
\put(1426,544){\rule[-0.175pt]{2.409pt}{0.350pt}}
\put(264,551){\rule[-0.175pt]{4.818pt}{0.350pt}}
\put(242,551){\makebox(0,0)[r]{$0.1$}}
\put(1416,551){\rule[-0.175pt]{4.818pt}{0.350pt}}
\put(264,575){\rule[-0.175pt]{2.409pt}{0.350pt}}
\put(1426,575){\rule[-0.175pt]{2.409pt}{0.350pt}}
\put(264,606){\rule[-0.175pt]{2.409pt}{0.350pt}}
\put(1426,606){\rule[-0.175pt]{2.409pt}{0.350pt}}
\put(264,622){\rule[-0.175pt]{2.409pt}{0.350pt}}
\put(1426,622){\rule[-0.175pt]{2.409pt}{0.350pt}}
\put(264,630){\rule[-0.175pt]{4.818pt}{0.350pt}}
\put(242,630){\makebox(0,0)[r]{$1$}}
\put(1416,630){\rule[-0.175pt]{4.818pt}{0.350pt}}
\put(264,653){\rule[-0.175pt]{2.409pt}{0.350pt}}
\put(1426,653){\rule[-0.175pt]{2.409pt}{0.350pt}}
\put(264,685){\rule[-0.175pt]{2.409pt}{0.350pt}}
\put(1426,685){\rule[-0.175pt]{2.409pt}{0.350pt}}
\put(264,701){\rule[-0.175pt]{2.409pt}{0.350pt}}
\put(1426,701){\rule[-0.175pt]{2.409pt}{0.350pt}}
\put(264,708){\rule[-0.175pt]{4.818pt}{0.350pt}}
\put(242,708){\makebox(0,0)[r]{$10$}}
\put(1416,708){\rule[-0.175pt]{4.818pt}{0.350pt}}
\put(264,732){\rule[-0.175pt]{2.409pt}{0.350pt}}
\put(1426,732){\rule[-0.175pt]{2.409pt}{0.350pt}}
\put(264,763){\rule[-0.175pt]{2.409pt}{0.350pt}}
\put(1426,763){\rule[-0.175pt]{2.409pt}{0.350pt}}
\put(264,779){\rule[-0.175pt]{2.409pt}{0.350pt}}
\put(1426,779){\rule[-0.175pt]{2.409pt}{0.350pt}}
\put(264,787){\rule[-0.175pt]{4.818pt}{0.350pt}}
\put(242,787){\makebox(0,0)[r]{$100$}}
\put(1416,787){\rule[-0.175pt]{4.818pt}{0.350pt}}
\put(264,158){\rule[-0.175pt]{0.350pt}{4.818pt}}
\put(264,113){\makebox(0,0){$0.01$}}
\put(264,767){\rule[-0.175pt]{0.350pt}{4.818pt}}
\put(417,158){\rule[-0.175pt]{0.350pt}{2.409pt}}
\put(417,777){\rule[-0.175pt]{0.350pt}{2.409pt}}
\put(507,158){\rule[-0.175pt]{0.350pt}{2.409pt}}
\put(507,777){\rule[-0.175pt]{0.350pt}{2.409pt}}
\put(571,158){\rule[-0.175pt]{0.350pt}{2.409pt}}
\put(571,777){\rule[-0.175pt]{0.350pt}{2.409pt}}
\put(620,158){\rule[-0.175pt]{0.350pt}{2.409pt}}
\put(620,777){\rule[-0.175pt]{0.350pt}{2.409pt}}
\put(660,158){\rule[-0.175pt]{0.350pt}{2.409pt}}
\put(660,777){\rule[-0.175pt]{0.350pt}{2.409pt}}
\put(694,158){\rule[-0.175pt]{0.350pt}{2.409pt}}
\put(694,777){\rule[-0.175pt]{0.350pt}{2.409pt}}
\put(724,158){\rule[-0.175pt]{0.350pt}{2.409pt}}
\put(724,777){\rule[-0.175pt]{0.350pt}{2.409pt}}
\put(750,158){\rule[-0.175pt]{0.350pt}{2.409pt}}
\put(750,777){\rule[-0.175pt]{0.350pt}{2.409pt}}
\put(773,158){\rule[-0.175pt]{0.350pt}{4.818pt}}
\put(773,113){\makebox(0,0){$0.1$}}
\put(773,767){\rule[-0.175pt]{0.350pt}{4.818pt}}
\put(927,158){\rule[-0.175pt]{0.350pt}{2.409pt}}
\put(927,777){\rule[-0.175pt]{0.350pt}{2.409pt}}
\put(1016,158){\rule[-0.175pt]{0.350pt}{2.409pt}}
\put(1016,777){\rule[-0.175pt]{0.350pt}{2.409pt}}
\put(1080,158){\rule[-0.175pt]{0.350pt}{2.409pt}}
\put(1080,777){\rule[-0.175pt]{0.350pt}{2.409pt}}
\put(1129,158){\rule[-0.175pt]{0.350pt}{2.409pt}}
\put(1129,777){\rule[-0.175pt]{0.350pt}{2.409pt}}
\put(1170,158){\rule[-0.175pt]{0.350pt}{2.409pt}}
\put(1170,777){\rule[-0.175pt]{0.350pt}{2.409pt}}
\put(1204,158){\rule[-0.175pt]{0.350pt}{2.409pt}}
\put(1204,777){\rule[-0.175pt]{0.350pt}{2.409pt}}
\put(1233,158){\rule[-0.175pt]{0.350pt}{2.409pt}}
\put(1233,777){\rule[-0.175pt]{0.350pt}{2.409pt}}
\put(1259,158){\rule[-0.175pt]{0.350pt}{2.409pt}}
\put(1259,777){\rule[-0.175pt]{0.350pt}{2.409pt}}
\put(1283,158){\rule[-0.175pt]{0.350pt}{4.818pt}}
\put(1283,113){\makebox(0,0){$1$}}
\put(1283,767){\rule[-0.175pt]{0.350pt}{4.818pt}}
\put(1436,158){\rule[-0.175pt]{0.350pt}{2.409pt}}
\put(1436,777){\rule[-0.175pt]{0.350pt}{2.409pt}}
\put(264,158){\rule[-0.175pt]{282.335pt}{0.350pt}}
\put(1436,158){\rule[-0.175pt]{0.350pt}{151.526pt}}
\put(264,787){\rule[-0.175pt]{282.335pt}{0.350pt}}
\put(45,472){\makebox(0,0)[l]{\shortstack{$H$}}}
\put(850,68){\makebox(0,0){${k \over k^{*}}$}}
\put(264,158){\rule[-0.175pt]{0.350pt}{151.526pt}}
\sbox{\plotpoint}{\rule[-0.500pt]{1.000pt}{1.000pt}}%
\put(264,758){\rule[-0.500pt]{1.474pt}{1.000pt}}
\put(270,757){\rule[-0.500pt]{1.474pt}{1.000pt}}
\put(276,756){\rule[-0.500pt]{1.474pt}{1.000pt}}
\put(282,755){\rule[-0.500pt]{1.474pt}{1.000pt}}
\put(288,754){\rule[-0.500pt]{1.474pt}{1.000pt}}
\put(294,753){\rule[-0.500pt]{1.474pt}{1.000pt}}
\put(300,752){\rule[-0.500pt]{1.474pt}{1.000pt}}
\put(306,751){\rule[-0.500pt]{1.474pt}{1.000pt}}
\put(312,750){\rule[-0.500pt]{1.474pt}{1.000pt}}
\put(319,749){\rule[-0.500pt]{1.474pt}{1.000pt}}
\put(325,748){\rule[-0.500pt]{1.474pt}{1.000pt}}
\put(331,747){\rule[-0.500pt]{1.474pt}{1.000pt}}
\put(337,746){\rule[-0.500pt]{1.474pt}{1.000pt}}
\put(343,745){\rule[-0.500pt]{1.474pt}{1.000pt}}
\put(349,744){\rule[-0.500pt]{1.474pt}{1.000pt}}
\put(355,743){\rule[-0.500pt]{1.474pt}{1.000pt}}
\put(361,742){\rule[-0.500pt]{1.474pt}{1.000pt}}
\put(368,741){\rule[-0.500pt]{1.474pt}{1.000pt}}
\put(374,740){\rule[-0.500pt]{1.474pt}{1.000pt}}
\put(380,739){\rule[-0.500pt]{1.474pt}{1.000pt}}
\put(386,738){\rule[-0.500pt]{1.474pt}{1.000pt}}
\put(392,737){\rule[-0.500pt]{1.474pt}{1.000pt}}
\put(398,736){\rule[-0.500pt]{1.474pt}{1.000pt}}
\put(404,735){\rule[-0.500pt]{1.474pt}{1.000pt}}
\put(410,734){\rule[-0.500pt]{1.474pt}{1.000pt}}
\put(416,733){\rule[-0.500pt]{1.549pt}{1.000pt}}
\put(423,732){\rule[-0.500pt]{1.549pt}{1.000pt}}
\put(429,731){\rule[-0.500pt]{1.549pt}{1.000pt}}
\put(436,730){\rule[-0.500pt]{1.549pt}{1.000pt}}
\put(442,729){\rule[-0.500pt]{1.549pt}{1.000pt}}
\put(449,728){\rule[-0.500pt]{1.549pt}{1.000pt}}
\put(455,727){\rule[-0.500pt]{1.549pt}{1.000pt}}
\put(461,726){\rule[-0.500pt]{1.549pt}{1.000pt}}
\put(468,725){\rule[-0.500pt]{1.549pt}{1.000pt}}
\put(474,724){\rule[-0.500pt]{1.549pt}{1.000pt}}
\put(481,723){\rule[-0.500pt]{1.549pt}{1.000pt}}
\put(487,722){\rule[-0.500pt]{1.549pt}{1.000pt}}
\put(494,721){\rule[-0.500pt]{1.549pt}{1.000pt}}
\put(500,720){\rule[-0.500pt]{1.549pt}{1.000pt}}
\put(506,719){\rule[-0.500pt]{1.542pt}{1.000pt}}
\put(513,718){\rule[-0.500pt]{1.542pt}{1.000pt}}
\put(519,717){\rule[-0.500pt]{1.542pt}{1.000pt}}
\put(526,716){\rule[-0.500pt]{1.542pt}{1.000pt}}
\put(532,715){\rule[-0.500pt]{1.542pt}{1.000pt}}
\put(539,714){\rule[-0.500pt]{1.542pt}{1.000pt}}
\put(545,713){\rule[-0.500pt]{1.542pt}{1.000pt}}
\put(551,712){\rule[-0.500pt]{1.542pt}{1.000pt}}
\put(558,711){\rule[-0.500pt]{1.542pt}{1.000pt}}
\put(564,710){\rule[-0.500pt]{1.542pt}{1.000pt}}
\put(571,709){\rule[-0.500pt]{1.476pt}{1.000pt}}
\put(577,708){\rule[-0.500pt]{1.476pt}{1.000pt}}
\put(583,707){\rule[-0.500pt]{1.476pt}{1.000pt}}
\put(589,706){\rule[-0.500pt]{1.476pt}{1.000pt}}
\put(595,705){\rule[-0.500pt]{1.476pt}{1.000pt}}
\put(601,704){\rule[-0.500pt]{1.476pt}{1.000pt}}
\put(607,703){\rule[-0.500pt]{1.476pt}{1.000pt}}
\put(613,702){\rule[-0.500pt]{1.476pt}{1.000pt}}
\put(620,701){\rule[-0.500pt]{1.606pt}{1.000pt}}
\put(626,700){\rule[-0.500pt]{1.606pt}{1.000pt}}
\put(633,699){\rule[-0.500pt]{1.606pt}{1.000pt}}
\put(640,698){\rule[-0.500pt]{1.606pt}{1.000pt}}
\put(646,697){\rule[-0.500pt]{1.606pt}{1.000pt}}
\put(653,696){\rule[-0.500pt]{1.606pt}{1.000pt}}
\put(660,695){\rule[-0.500pt]{1.365pt}{1.000pt}}
\put(665,694){\rule[-0.500pt]{1.365pt}{1.000pt}}
\put(671,693){\rule[-0.500pt]{1.365pt}{1.000pt}}
\put(677,692){\rule[-0.500pt]{1.365pt}{1.000pt}}
\put(682,691){\rule[-0.500pt]{1.365pt}{1.000pt}}
\put(688,690){\rule[-0.500pt]{1.365pt}{1.000pt}}
\put(694,689){\rule[-0.500pt]{1.807pt}{1.000pt}}
\put(701,688){\rule[-0.500pt]{1.807pt}{1.000pt}}
\put(709,687){\rule[-0.500pt]{1.807pt}{1.000pt}}
\put(716,686){\rule[-0.500pt]{1.807pt}{1.000pt}}
\put(724,685){\rule[-0.500pt]{1.566pt}{1.000pt}}
\put(730,684){\rule[-0.500pt]{1.566pt}{1.000pt}}
\put(737,683){\rule[-0.500pt]{1.566pt}{1.000pt}}
\put(743,682){\rule[-0.500pt]{1.566pt}{1.000pt}}
\put(750,681){\rule[-0.500pt]{1.385pt}{1.000pt}}
\put(755,680){\rule[-0.500pt]{1.385pt}{1.000pt}}
\put(761,679){\rule[-0.500pt]{1.385pt}{1.000pt}}
\put(767,678){\rule[-0.500pt]{1.385pt}{1.000pt}}
\put(773,677){\rule[-0.500pt]{1.265pt}{1.000pt}}
\put(778,676){\rule[-0.500pt]{1.265pt}{1.000pt}}
\put(783,675){\rule[-0.500pt]{1.265pt}{1.000pt}}
\put(788,674){\rule[-0.500pt]{1.265pt}{1.000pt}}
\put(794,673){\rule[-0.500pt]{1.606pt}{1.000pt}}
\put(800,672){\rule[-0.500pt]{1.606pt}{1.000pt}}
\put(807,671){\rule[-0.500pt]{1.606pt}{1.000pt}}
\put(814,670){\rule[-0.500pt]{2.048pt}{1.000pt}}
\put(822,669){\rule[-0.500pt]{2.048pt}{1.000pt}}
\put(831,668){\rule[-0.500pt]{1.365pt}{1.000pt}}
\put(836,667){\rule[-0.500pt]{1.365pt}{1.000pt}}
\put(842,666){\rule[-0.500pt]{1.365pt}{1.000pt}}
\put(848,665){\rule[-0.500pt]{1.204pt}{1.000pt}}
\put(853,664){\rule[-0.500pt]{1.204pt}{1.000pt}}
\put(858,663){\rule[-0.500pt]{1.204pt}{1.000pt}}
\put(863,662){\rule[-0.500pt]{1.686pt}{1.000pt}}
\put(870,661){\rule[-0.500pt]{1.686pt}{1.000pt}}
\put(877,660){\rule[-0.500pt]{1.686pt}{1.000pt}}
\put(884,659){\rule[-0.500pt]{1.686pt}{1.000pt}}
\put(891,658){\rule[-0.500pt]{1.445pt}{1.000pt}}
\put(897,657){\rule[-0.500pt]{1.445pt}{1.000pt}}
\put(903,656){\rule[-0.500pt]{1.445pt}{1.000pt}}
\put(909,655){\rule[-0.500pt]{1.445pt}{1.000pt}}
\put(915,654){\rule[-0.500pt]{1.445pt}{1.000pt}}
\put(921,653){\rule[-0.500pt]{1.445pt}{1.000pt}}
\put(927,652){\rule[-0.500pt]{2.409pt}{1.000pt}}
\put(937,651){\rule[-0.500pt]{1.325pt}{1.000pt}}
\put(942,650){\rule[-0.500pt]{1.325pt}{1.000pt}}
\put(948,649){\rule[-0.500pt]{1.204pt}{1.000pt}}
\put(953,648){\rule[-0.500pt]{1.204pt}{1.000pt}}
\put(958,647){\rule[-0.500pt]{2.168pt}{1.000pt}}
\put(967,646){\rule[-0.500pt]{1.084pt}{1.000pt}}
\put(971,645){\rule[-0.500pt]{1.084pt}{1.000pt}}
\put(976,644){\rule[-0.500pt]{2.168pt}{1.000pt}}
\put(985,643){\usebox{\plotpoint}}
\put(989,642){\usebox{\plotpoint}}
\put(993,641){\rule[-0.500pt]{1.927pt}{1.000pt}}
\put(1001,640){\rule[-0.500pt]{1.927pt}{1.000pt}}
\put(1009,639){\rule[-0.500pt]{1.686pt}{1.000pt}}
\put(1016,638){\usebox{\plotpoint}}
\put(1020,637){\usebox{\plotpoint}}
\put(1024,636){\rule[-0.500pt]{1.686pt}{1.000pt}}
\put(1031,635){\rule[-0.500pt]{1.445pt}{1.000pt}}
\put(1037,634){\rule[-0.500pt]{1.686pt}{1.000pt}}
\put(1044,633){\rule[-0.500pt]{1.445pt}{1.000pt}}
\put(1050,632){\rule[-0.500pt]{1.686pt}{1.000pt}}
\put(1057,631){\rule[-0.500pt]{1.445pt}{1.000pt}}
\put(1063,630){\rule[-0.500pt]{1.445pt}{1.000pt}}
\put(1069,629){\rule[-0.500pt]{1.204pt}{1.000pt}}
\put(1074,628){\rule[-0.500pt]{1.445pt}{1.000pt}}
\put(1080,627){\rule[-0.500pt]{1.204pt}{1.000pt}}
\put(1085,626){\rule[-0.500pt]{1.445pt}{1.000pt}}
\put(1091,625){\rule[-0.500pt]{1.204pt}{1.000pt}}
\put(1096,624){\rule[-0.500pt]{1.204pt}{1.000pt}}
\put(1101,623){\rule[-0.500pt]{1.204pt}{1.000pt}}
\put(1106,622){\rule[-0.500pt]{1.204pt}{1.000pt}}
\put(1111,621){\rule[-0.500pt]{1.204pt}{1.000pt}}
\put(1116,620){\usebox{\plotpoint}}
\put(1120,619){\rule[-0.500pt]{1.204pt}{1.000pt}}
\put(1125,618){\usebox{\plotpoint}}
\put(1129,617){\rule[-0.500pt]{1.204pt}{1.000pt}}
\put(1134,616){\usebox{\plotpoint}}
\put(1138,615){\usebox{\plotpoint}}
\put(1142,614){\usebox{\plotpoint}}
\put(1146,613){\usebox{\plotpoint}}
\put(1150,612){\usebox{\plotpoint}}
\put(1154,611){\rule[-0.500pt]{1.927pt}{1.000pt}}
\put(1162,610){\usebox{\plotpoint}}
\put(1166,609){\usebox{\plotpoint}}
\put(1170,608){\usebox{\plotpoint}}
\put(1173,607){\usebox{\plotpoint}}
\put(1177,606){\usebox{\plotpoint}}
\put(1178,605){\usebox{\plotpoint}}
\put(1180,604){\usebox{\plotpoint}}
\put(1184,603){\usebox{\plotpoint}}
\put(1187,602){\usebox{\plotpoint}}
\put(1191,601){\usebox{\plotpoint}}
\put(1194,600){\usebox{\plotpoint}}
\put(1197,599){\usebox{\plotpoint}}
\put(1201,598){\usebox{\plotpoint}}
\put(1204,597){\usebox{\plotpoint}}
\put(1205,596){\usebox{\plotpoint}}
\put(1207,595){\usebox{\plotpoint}}
\put(1210,594){\usebox{\plotpoint}}
\put(1213,593){\usebox{\plotpoint}}
\put(1214,592){\usebox{\plotpoint}}
\put(1216,591){\usebox{\plotpoint}}
\put(1219,590){\usebox{\plotpoint}}
\put(1220,589){\usebox{\plotpoint}}
\put(1222,588){\usebox{\plotpoint}}
\put(1225,587){\usebox{\plotpoint}}
\put(1226,586){\usebox{\plotpoint}}
\put(1228,585){\usebox{\plotpoint}}
\put(1231,584){\usebox{\plotpoint}}
\put(1232,583){\usebox{\plotpoint}}
\put(1233,582){\usebox{\plotpoint}}
\put(1234,581){\usebox{\plotpoint}}
\put(1236,580){\usebox{\plotpoint}}
\put(1237,579){\usebox{\plotpoint}}
\put(1239,578){\usebox{\plotpoint}}
\put(1240,577){\usebox{\plotpoint}}
\put(1241,576){\usebox{\plotpoint}}
\put(1242,575){\usebox{\plotpoint}}
\put(1244,574){\usebox{\plotpoint}}
\put(1245,573){\usebox{\plotpoint}}
\put(1247,570){\usebox{\plotpoint}}
\put(1248,569){\usebox{\plotpoint}}
\put(1249,569){\usebox{\plotpoint}}
\put(1250,568){\usebox{\plotpoint}}
\put(1252,565){\usebox{\plotpoint}}
\put(1253,564){\usebox{\plotpoint}}
\put(1254,564){\usebox{\plotpoint}}
\put(1255,563){\usebox{\plotpoint}}
\put(1256,562){\usebox{\plotpoint}}
\put(1257,559){\usebox{\plotpoint}}
\put(1258,558){\usebox{\plotpoint}}
\put(1259,556){\usebox{\plotpoint}}
\put(1260,555){\usebox{\plotpoint}}
\put(1261,554){\usebox{\plotpoint}}
\put(1262,552){\usebox{\plotpoint}}
\put(1263,550){\usebox{\plotpoint}}
\put(1264,548){\usebox{\plotpoint}}
\put(1265,546){\usebox{\plotpoint}}
\put(1266,545){\usebox{\plotpoint}}
\put(1267,542){\usebox{\plotpoint}}
\put(1268,540){\usebox{\plotpoint}}
\put(1269,537){\usebox{\plotpoint}}
\put(1270,534){\usebox{\plotpoint}}
\put(1271,531){\usebox{\plotpoint}}
\put(1272,528){\usebox{\plotpoint}}
\put(1273,526){\usebox{\plotpoint}}
\put(1274,521){\rule[-0.500pt]{1.000pt}{1.204pt}}
\put(1275,516){\rule[-0.500pt]{1.000pt}{1.204pt}}
\put(1276,509){\rule[-0.500pt]{1.000pt}{1.566pt}}
\put(1277,503){\rule[-0.500pt]{1.000pt}{1.566pt}}
\put(1278,491){\rule[-0.500pt]{1.000pt}{2.891pt}}
\put(1279,479){\rule[-0.500pt]{1.000pt}{2.891pt}}
\put(1280,392){\rule[-0.500pt]{1.000pt}{20.878pt}}
\put(1281,305){\rule[-0.500pt]{1.000pt}{20.878pt}}
\put(1282,219){\rule[-0.500pt]{1.000pt}{20.878pt}}
\put(1283,219){\usebox{\plotpoint}}
\sbox{\plotpoint}{\rule[-0.350pt]{0.700pt}{0.700pt}}%
\put(264,769){\rule[-0.350pt]{1.603pt}{0.700pt}}
\put(270,768){\rule[-0.350pt]{1.603pt}{0.700pt}}
\put(277,767){\rule[-0.350pt]{1.603pt}{0.700pt}}
\put(283,766){\rule[-0.350pt]{1.603pt}{0.700pt}}
\put(290,765){\rule[-0.350pt]{1.603pt}{0.700pt}}
\put(297,764){\rule[-0.350pt]{1.603pt}{0.700pt}}
\put(303,763){\rule[-0.350pt]{1.603pt}{0.700pt}}
\put(310,762){\rule[-0.350pt]{1.603pt}{0.700pt}}
\put(317,761){\rule[-0.350pt]{1.603pt}{0.700pt}}
\put(323,760){\rule[-0.350pt]{1.603pt}{0.700pt}}
\put(330,759){\rule[-0.350pt]{1.603pt}{0.700pt}}
\put(337,758){\rule[-0.350pt]{1.603pt}{0.700pt}}
\put(343,757){\rule[-0.350pt]{1.603pt}{0.700pt}}
\put(350,756){\rule[-0.350pt]{1.603pt}{0.700pt}}
\put(357,755){\rule[-0.350pt]{1.603pt}{0.700pt}}
\put(363,754){\rule[-0.350pt]{1.603pt}{0.700pt}}
\put(370,753){\rule[-0.350pt]{1.603pt}{0.700pt}}
\put(377,752){\rule[-0.350pt]{1.603pt}{0.700pt}}
\put(383,751){\rule[-0.350pt]{1.603pt}{0.700pt}}
\put(390,750){\rule[-0.350pt]{1.603pt}{0.700pt}}
\put(397,749){\rule[-0.350pt]{1.603pt}{0.700pt}}
\put(403,748){\rule[-0.350pt]{1.603pt}{0.700pt}}
\put(410,747){\rule[-0.350pt]{1.603pt}{0.700pt}}
\put(416,746){\rule[-0.350pt]{1.549pt}{0.700pt}}
\put(423,745){\rule[-0.350pt]{1.549pt}{0.700pt}}
\put(429,744){\rule[-0.350pt]{1.549pt}{0.700pt}}
\put(436,743){\rule[-0.350pt]{1.549pt}{0.700pt}}
\put(442,742){\rule[-0.350pt]{1.549pt}{0.700pt}}
\put(449,741){\rule[-0.350pt]{1.549pt}{0.700pt}}
\put(455,740){\rule[-0.350pt]{1.549pt}{0.700pt}}
\put(461,739){\rule[-0.350pt]{1.549pt}{0.700pt}}
\put(468,738){\rule[-0.350pt]{1.549pt}{0.700pt}}
\put(474,737){\rule[-0.350pt]{1.549pt}{0.700pt}}
\put(481,736){\rule[-0.350pt]{1.549pt}{0.700pt}}
\put(487,735){\rule[-0.350pt]{1.549pt}{0.700pt}}
\put(494,734){\rule[-0.350pt]{1.549pt}{0.700pt}}
\put(500,733){\rule[-0.350pt]{1.549pt}{0.700pt}}
\put(506,732){\rule[-0.350pt]{1.713pt}{0.700pt}}
\put(514,731){\rule[-0.350pt]{1.713pt}{0.700pt}}
\put(521,730){\rule[-0.350pt]{1.713pt}{0.700pt}}
\put(528,729){\rule[-0.350pt]{1.713pt}{0.700pt}}
\put(535,728){\rule[-0.350pt]{1.713pt}{0.700pt}}
\put(542,727){\rule[-0.350pt]{1.713pt}{0.700pt}}
\put(549,726){\rule[-0.350pt]{1.713pt}{0.700pt}}
\put(556,725){\rule[-0.350pt]{1.713pt}{0.700pt}}
\put(563,724){\rule[-0.350pt]{1.713pt}{0.700pt}}
\put(570,723){\rule[-0.350pt]{1.476pt}{0.700pt}}
\put(577,722){\rule[-0.350pt]{1.476pt}{0.700pt}}
\put(583,721){\rule[-0.350pt]{1.476pt}{0.700pt}}
\put(589,720){\rule[-0.350pt]{1.476pt}{0.700pt}}
\put(595,719){\rule[-0.350pt]{1.476pt}{0.700pt}}
\put(601,718){\rule[-0.350pt]{1.476pt}{0.700pt}}
\put(607,717){\rule[-0.350pt]{1.476pt}{0.700pt}}
\put(613,716){\rule[-0.350pt]{1.476pt}{0.700pt}}
\put(620,715){\rule[-0.350pt]{1.606pt}{0.700pt}}
\put(626,714){\rule[-0.350pt]{1.606pt}{0.700pt}}
\put(633,713){\rule[-0.350pt]{1.606pt}{0.700pt}}
\put(640,712){\rule[-0.350pt]{1.606pt}{0.700pt}}
\put(646,711){\rule[-0.350pt]{1.606pt}{0.700pt}}
\put(653,710){\rule[-0.350pt]{1.606pt}{0.700pt}}
\put(660,709){\rule[-0.350pt]{1.638pt}{0.700pt}}
\put(666,708){\rule[-0.350pt]{1.638pt}{0.700pt}}
\put(673,707){\rule[-0.350pt]{1.638pt}{0.700pt}}
\put(680,706){\rule[-0.350pt]{1.638pt}{0.700pt}}
\put(687,705){\rule[-0.350pt]{1.638pt}{0.700pt}}
\put(693,704){\rule[-0.350pt]{1.807pt}{0.700pt}}
\put(701,703){\rule[-0.350pt]{1.807pt}{0.700pt}}
\put(709,702){\rule[-0.350pt]{1.807pt}{0.700pt}}
\put(716,701){\rule[-0.350pt]{1.807pt}{0.700pt}}
\put(724,700){\rule[-0.350pt]{1.566pt}{0.700pt}}
\put(730,699){\rule[-0.350pt]{1.566pt}{0.700pt}}
\put(737,698){\rule[-0.350pt]{1.566pt}{0.700pt}}
\put(743,697){\rule[-0.350pt]{1.566pt}{0.700pt}}
\put(750,696){\rule[-0.350pt]{1.847pt}{0.700pt}}
\put(757,695){\rule[-0.350pt]{1.847pt}{0.700pt}}
\put(765,694){\rule[-0.350pt]{1.847pt}{0.700pt}}
\put(773,693){\rule[-0.350pt]{1.686pt}{0.700pt}}
\put(780,692){\rule[-0.350pt]{1.686pt}{0.700pt}}
\put(787,691){\rule[-0.350pt]{1.686pt}{0.700pt}}
\put(794,690){\rule[-0.350pt]{1.606pt}{0.700pt}}
\put(800,689){\rule[-0.350pt]{1.606pt}{0.700pt}}
\put(807,688){\rule[-0.350pt]{1.606pt}{0.700pt}}
\put(814,687){\rule[-0.350pt]{1.365pt}{0.700pt}}
\put(819,686){\rule[-0.350pt]{1.365pt}{0.700pt}}
\put(825,685){\rule[-0.350pt]{1.365pt}{0.700pt}}
\put(831,684){\rule[-0.350pt]{2.048pt}{0.700pt}}
\put(839,683){\rule[-0.350pt]{2.048pt}{0.700pt}}
\put(848,682){\rule[-0.350pt]{1.807pt}{0.700pt}}
\put(855,681){\rule[-0.350pt]{1.807pt}{0.700pt}}
\put(863,680){\rule[-0.350pt]{1.686pt}{0.700pt}}
\put(870,679){\rule[-0.350pt]{1.686pt}{0.700pt}}
\put(877,678){\rule[-0.350pt]{1.686pt}{0.700pt}}
\put(884,677){\rule[-0.350pt]{1.686pt}{0.700pt}}
\put(891,676){\rule[-0.350pt]{1.445pt}{0.700pt}}
\put(897,675){\rule[-0.350pt]{1.445pt}{0.700pt}}
\put(903,674){\rule[-0.350pt]{1.445pt}{0.700pt}}
\put(909,673){\rule[-0.350pt]{1.445pt}{0.700pt}}
\put(915,672){\rule[-0.350pt]{2.891pt}{0.700pt}}
\put(927,671){\rule[-0.350pt]{1.204pt}{0.700pt}}
\put(932,670){\rule[-0.350pt]{1.204pt}{0.700pt}}
\put(937,669){\rule[-0.350pt]{1.325pt}{0.700pt}}
\put(942,668){\rule[-0.350pt]{1.325pt}{0.700pt}}
\put(948,667){\rule[-0.350pt]{2.409pt}{0.700pt}}
\put(958,666){\rule[-0.350pt]{2.168pt}{0.700pt}}
\put(967,665){\rule[-0.350pt]{1.084pt}{0.700pt}}
\put(971,664){\rule[-0.350pt]{1.084pt}{0.700pt}}
\put(976,663){\rule[-0.350pt]{2.168pt}{0.700pt}}
\put(985,662){\rule[-0.350pt]{1.927pt}{0.700pt}}
\put(993,661){\rule[-0.350pt]{0.964pt}{0.700pt}}
\put(997,660){\rule[-0.350pt]{0.964pt}{0.700pt}}
\put(1001,659){\rule[-0.350pt]{1.927pt}{0.700pt}}
\put(1009,658){\rule[-0.350pt]{1.686pt}{0.700pt}}
\put(1016,657){\rule[-0.350pt]{1.927pt}{0.700pt}}
\put(1024,656){\rule[-0.350pt]{1.686pt}{0.700pt}}
\put(1031,655){\rule[-0.350pt]{1.445pt}{0.700pt}}
\put(1037,654){\rule[-0.350pt]{1.686pt}{0.700pt}}
\put(1044,653){\rule[-0.350pt]{1.445pt}{0.700pt}}
\put(1050,652){\rule[-0.350pt]{0.843pt}{0.700pt}}
\put(1053,651){\rule[-0.350pt]{0.843pt}{0.700pt}}
\put(1057,650){\rule[-0.350pt]{1.445pt}{0.700pt}}
\put(1063,649){\rule[-0.350pt]{1.445pt}{0.700pt}}
\put(1069,648){\rule[-0.350pt]{1.204pt}{0.700pt}}
\put(1074,647){\rule[-0.350pt]{1.445pt}{0.700pt}}
\put(1080,646){\rule[-0.350pt]{2.650pt}{0.700pt}}
\put(1091,645){\rule[-0.350pt]{1.204pt}{0.700pt}}
\put(1096,644){\rule[-0.350pt]{1.204pt}{0.700pt}}
\put(1101,643){\rule[-0.350pt]{1.204pt}{0.700pt}}
\put(1106,642){\rule[-0.350pt]{1.204pt}{0.700pt}}
\put(1111,641){\rule[-0.350pt]{1.204pt}{0.700pt}}
\put(1116,640){\rule[-0.350pt]{0.964pt}{0.700pt}}
\put(1120,639){\rule[-0.350pt]{1.204pt}{0.700pt}}
\put(1125,638){\rule[-0.350pt]{0.964pt}{0.700pt}}
\put(1129,637){\rule[-0.350pt]{2.168pt}{0.700pt}}
\put(1138,636){\rule[-0.350pt]{0.964pt}{0.700pt}}
\put(1142,635){\rule[-0.350pt]{0.964pt}{0.700pt}}
\put(1146,634){\rule[-0.350pt]{0.964pt}{0.700pt}}
\put(1150,633){\rule[-0.350pt]{0.964pt}{0.700pt}}
\put(1154,632){\rule[-0.350pt]{0.964pt}{0.700pt}}
\put(1158,631){\rule[-0.350pt]{1.927pt}{0.700pt}}
\put(1166,630){\rule[-0.350pt]{0.964pt}{0.700pt}}
\put(1170,629){\rule[-0.350pt]{0.723pt}{0.700pt}}
\put(1173,628){\rule[-0.350pt]{0.964pt}{0.700pt}}
\put(1177,627){\rule[-0.350pt]{0.723pt}{0.700pt}}
\put(1180,626){\rule[-0.350pt]{1.686pt}{0.700pt}}
\put(1187,625){\rule[-0.350pt]{0.964pt}{0.700pt}}
\put(1191,624){\rule[-0.350pt]{0.723pt}{0.700pt}}
\put(1194,623){\rule[-0.350pt]{0.723pt}{0.700pt}}
\put(1197,622){\rule[-0.350pt]{0.964pt}{0.700pt}}
\put(1201,621){\rule[-0.350pt]{1.445pt}{0.700pt}}
\put(1207,620){\rule[-0.350pt]{0.723pt}{0.700pt}}
\put(1210,619){\rule[-0.350pt]{0.723pt}{0.700pt}}
\put(1213,618){\rule[-0.350pt]{0.723pt}{0.700pt}}
\put(1216,617){\rule[-0.350pt]{0.723pt}{0.700pt}}
\put(1219,616){\rule[-0.350pt]{1.445pt}{0.700pt}}
\put(1225,615){\rule[-0.350pt]{0.723pt}{0.700pt}}
\put(1228,614){\rule[-0.350pt]{0.723pt}{0.700pt}}
\put(1231,613){\usebox{\plotpoint}}
\put(1233,612){\rule[-0.350pt]{0.723pt}{0.700pt}}
\put(1236,611){\rule[-0.350pt]{1.204pt}{0.700pt}}
\put(1241,610){\rule[-0.350pt]{0.723pt}{0.700pt}}
\put(1244,609){\rule[-0.350pt]{0.723pt}{0.700pt}}
\put(1247,608){\usebox{\plotpoint}}
\put(1249,607){\rule[-0.350pt]{0.723pt}{0.700pt}}
\put(1252,606){\usebox{\plotpoint}}
\put(1254,605){\rule[-0.350pt]{1.204pt}{0.700pt}}
\put(1259,604){\rule[-0.350pt]{0.723pt}{0.700pt}}
\put(1262,603){\usebox{\plotpoint}}
\put(1264,602){\rule[-0.350pt]{0.723pt}{0.700pt}}
\put(1267,601){\usebox{\plotpoint}}
\put(1269,600){\usebox{\plotpoint}}
\put(1271,599){\rule[-0.350pt]{0.723pt}{0.700pt}}
\put(1274,598){\usebox{\plotpoint}}
\put(1276,597){\usebox{\plotpoint}}
\put(1278,596){\rule[-0.350pt]{1.204pt}{0.700pt}}
\put(1283,595){\usebox{\plotpoint}}
\put(1285,594){\usebox{\plotpoint}}
\put(1287,593){\usebox{\plotpoint}}
\put(1289,592){\usebox{\plotpoint}}
\put(1291,591){\usebox{\plotpoint}}
\put(1293,590){\rule[-0.350pt]{0.723pt}{0.700pt}}
\put(1296,589){\usebox{\plotpoint}}
\put(1298,588){\usebox{\plotpoint}}
\put(1300,587){\usebox{\plotpoint}}
\put(1302,586){\usebox{\plotpoint}}
\put(1304,585){\usebox{\plotpoint}}
\put(1306,584){\usebox{\plotpoint}}
\put(1307,583){\usebox{\plotpoint}}
\put(1308,582){\usebox{\plotpoint}}
\put(1310,581){\usebox{\plotpoint}}
\put(1312,580){\usebox{\plotpoint}}
\put(1314,579){\usebox{\plotpoint}}
\put(1316,578){\usebox{\plotpoint}}
\put(1317,577){\usebox{\plotpoint}}
\put(1319,576){\usebox{\plotpoint}}
\put(1320,575){\usebox{\plotpoint}}
\put(1321,574){\usebox{\plotpoint}}
\put(1323,573){\usebox{\plotpoint}}
\put(1325,572){\usebox{\plotpoint}}
\put(1327,569){\usebox{\plotpoint}}
\put(1328,569){\usebox{\plotpoint}}
\put(1330,568){\usebox{\plotpoint}}
\put(1332,567){\usebox{\plotpoint}}
\put(1333,566){\usebox{\plotpoint}}
\put(1334,565){\usebox{\plotpoint}}
\put(1336,564){\usebox{\plotpoint}}
\put(1337,563){\usebox{\plotpoint}}
\put(1338,562){\usebox{\plotpoint}}
\put(1339,561){\usebox{\plotpoint}}
\put(1341,558){\usebox{\plotpoint}}
\put(1342,558){\usebox{\plotpoint}}
\put(1343,557){\usebox{\plotpoint}}
\put(1344,556){\usebox{\plotpoint}}
\put(1346,553){\usebox{\plotpoint}}
\put(1347,553){\usebox{\plotpoint}}
\put(1348,552){\usebox{\plotpoint}}
\put(1349,551){\usebox{\plotpoint}}
\put(1350,550){\usebox{\plotpoint}}
\put(1351,547){\usebox{\plotpoint}}
\put(1352,547){\usebox{\plotpoint}}
\put(1353,546){\usebox{\plotpoint}}
\put(1354,545){\usebox{\plotpoint}}
\put(1355,544){\usebox{\plotpoint}}
\put(1356,541){\usebox{\plotpoint}}
\put(1357,541){\usebox{\plotpoint}}
\put(1358,540){\usebox{\plotpoint}}
\put(1359,537){\usebox{\plotpoint}}
\put(1360,535){\usebox{\plotpoint}}
\put(1361,534){\usebox{\plotpoint}}
\put(1362,531){\rule[-0.350pt]{0.700pt}{0.723pt}}
\put(1363,531){\usebox{\plotpoint}}
\put(1364,530){\usebox{\plotpoint}}
\put(1365,526){\rule[-0.350pt]{0.700pt}{0.723pt}}
\put(1366,524){\usebox{\plotpoint}}
\put(1367,522){\usebox{\plotpoint}}
\put(1368,519){\rule[-0.350pt]{0.700pt}{0.723pt}}
\put(1369,517){\usebox{\plotpoint}}
\put(1370,515){\usebox{\plotpoint}}
\put(1371,510){\rule[-0.350pt]{0.700pt}{1.204pt}}
\put(1372,507){\usebox{\plotpoint}}
\put(1373,505){\usebox{\plotpoint}}
\put(1374,500){\rule[-0.350pt]{0.700pt}{1.204pt}}
\put(1375,496){\rule[-0.350pt]{0.700pt}{0.843pt}}
\put(1376,493){\rule[-0.350pt]{0.700pt}{0.843pt}}
\put(1377,485){\rule[-0.350pt]{0.700pt}{1.927pt}}
\put(1378,479){\rule[-0.350pt]{0.700pt}{1.325pt}}
\put(1379,474){\rule[-0.350pt]{0.700pt}{1.325pt}}
\put(1380,459){\rule[-0.350pt]{0.700pt}{3.613pt}}
\put(1381,432){\rule[-0.350pt]{0.700pt}{6.504pt}}
\put(1382,431){\usebox{\plotpoint}}
\put(1382,431){\usebox{\plotpoint}}
\put(1383,356){\rule[-0.350pt]{0.700pt}{17.827pt}}
\put(1384,221){\rule[-0.350pt]{0.700pt}{32.521pt}}
\end{picture}

%% file: fig2_inset.tex
\setlength{\unitlength}{0.240900pt}
\ifx\plotpoint\undefined\newsavebox{\plotpoint}\fi
\sbox{\plotpoint}{\rule[-0.175pt]{0.350pt}{0.350pt}}%
\begin{picture}(1500,900)(0,0)
\sbox{\plotpoint}{\rule[-0.175pt]{0.350pt}{0.350pt}}%
\put(264,697){\rule[-0.175pt]{282.335pt}{0.350pt}}
\put(264,158){\rule[-0.175pt]{4.818pt}{0.350pt}}
\put(242,158){\makebox(0,0)[r]{$-12$}}
\put(1416,158){\rule[-0.175pt]{4.818pt}{0.350pt}}
\put(264,248){\rule[-0.175pt]{4.818pt}{0.350pt}}
\put(242,248){\makebox(0,0)[r]{$-10$}}
\put(1416,248){\rule[-0.175pt]{4.818pt}{0.350pt}}
\put(264,338){\rule[-0.175pt]{4.818pt}{0.350pt}}
\put(242,338){\makebox(0,0)[r]{$-8$}}
\put(1416,338){\rule[-0.175pt]{4.818pt}{0.350pt}}
\put(264,428){\rule[-0.175pt]{4.818pt}{0.350pt}}
\put(242,428){\makebox(0,0)[r]{$-6$}}
\put(1416,428){\rule[-0.175pt]{4.818pt}{0.350pt}}
\put(264,517){\rule[-0.175pt]{4.818pt}{0.350pt}}
\put(242,517){\makebox(0,0)[r]{$-4$}}
\put(1416,517){\rule[-0.175pt]{4.818pt}{0.350pt}}
\put(264,607){\rule[-0.175pt]{4.818pt}{0.350pt}}
\put(242,607){\makebox(0,0)[r]{$-2$}}
\put(1416,607){\rule[-0.175pt]{4.818pt}{0.350pt}}
\put(264,697){\rule[-0.175pt]{4.818pt}{0.350pt}}
\put(242,697){\makebox(0,0)[r]{$0$}}
\put(1416,697){\rule[-0.175pt]{4.818pt}{0.350pt}}
\put(264,787){\rule[-0.175pt]{4.818pt}{0.350pt}}
\put(242,787){\makebox(0,0)[r]{$2$}}
\put(1416,787){\rule[-0.175pt]{4.818pt}{0.350pt}}
\put(264,158){\rule[-0.175pt]{0.350pt}{4.818pt}}
\put(264,113){\makebox(0,0){$0.1$}}
\put(264,767){\rule[-0.175pt]{0.350pt}{4.818pt}}
\put(394,158){\rule[-0.175pt]{0.350pt}{4.818pt}}
\put(394,113){\makebox(0,0){$0.2$}}
\put(394,767){\rule[-0.175pt]{0.350pt}{4.818pt}}
\put(524,158){\rule[-0.175pt]{0.350pt}{4.818pt}}
\put(524,113){\makebox(0,0){$0.3$}}
\put(524,767){\rule[-0.175pt]{0.350pt}{4.818pt}}
\put(655,158){\rule[-0.175pt]{0.350pt}{4.818pt}}
\put(655,113){\makebox(0,0){$0.4$}}
\put(655,767){\rule[-0.175pt]{0.350pt}{4.818pt}}
\put(785,158){\rule[-0.175pt]{0.350pt}{4.818pt}}
\put(785,113){\makebox(0,0){$0.5$}}
\put(785,767){\rule[-0.175pt]{0.350pt}{4.818pt}}
\put(915,158){\rule[-0.175pt]{0.350pt}{4.818pt}}
\put(915,113){\makebox(0,0){$0.6$}}
\put(915,767){\rule[-0.175pt]{0.350pt}{4.818pt}}
\put(1045,158){\rule[-0.175pt]{0.350pt}{4.818pt}}
\put(1045,113){\makebox(0,0){$0.7$}}
\put(1045,767){\rule[-0.175pt]{0.350pt}{4.818pt}}
\put(1176,158){\rule[-0.175pt]{0.350pt}{4.818pt}}
\put(1176,113){\makebox(0,0){$0.8$}}
\put(1176,767){\rule[-0.175pt]{0.350pt}{4.818pt}}
\put(1306,158){\rule[-0.175pt]{0.350pt}{4.818pt}}
\put(1306,113){\makebox(0,0){$0.9$}}
\put(1306,767){\rule[-0.175pt]{0.350pt}{4.818pt}}
\put(1436,158){\rule[-0.175pt]{0.350pt}{4.818pt}}
\put(1436,113){\makebox(0,0){$1$}}
\put(1436,767){\rule[-0.175pt]{0.350pt}{4.818pt}}
\put(264,158){\rule[-0.175pt]{282.335pt}{0.350pt}}
\put(1436,158){\rule[-0.175pt]{0.350pt}{151.526pt}}
\put(264,787){\rule[-0.175pt]{282.335pt}{0.350pt}}
\put(45,472){\makebox(0,0)[l]{\shortstack{$H$}}}
\put(850,68){\makebox(0,0){${k \over k^{*}}$}}
\put(277,643){\makebox(0,0)[l]{$z1$}}
\put(277,765){\makebox(0,0)[l]{$z2$}}
\put(277,715){\makebox(0,0)[l]{$z3$}}
\put(277,428){\makebox(0,0)[l]{$z4$}}
\put(264,158){\rule[-0.175pt]{0.350pt}{151.526pt}}
\sbox{\plotpoint}{\rule[-0.500pt]{1.000pt}{1.000pt}}%
\put(264,609){\usebox{\plotpoint}}
\put(264,609){\rule[-0.500pt]{1.566pt}{1.000pt}}
\put(270,610){\rule[-0.500pt]{1.566pt}{1.000pt}}
\put(277,611){\rule[-0.500pt]{3.132pt}{1.000pt}}
\put(290,612){\rule[-0.500pt]{3.132pt}{1.000pt}}
\put(303,613){\rule[-0.500pt]{3.132pt}{1.000pt}}
\put(316,614){\rule[-0.500pt]{3.132pt}{1.000pt}}
\put(329,615){\rule[-0.500pt]{3.132pt}{1.000pt}}
\put(342,616){\rule[-0.500pt]{3.132pt}{1.000pt}}
\put(355,617){\rule[-0.500pt]{3.132pt}{1.000pt}}
\put(368,618){\rule[-0.500pt]{6.263pt}{1.000pt}}
\put(394,619){\rule[-0.500pt]{3.132pt}{1.000pt}}
\put(407,620){\rule[-0.500pt]{3.132pt}{1.000pt}}
\put(420,621){\rule[-0.500pt]{6.263pt}{1.000pt}}
\put(446,622){\rule[-0.500pt]{3.132pt}{1.000pt}}
\put(459,623){\rule[-0.500pt]{6.263pt}{1.000pt}}
\put(485,624){\rule[-0.500pt]{3.132pt}{1.000pt}}
\put(498,625){\rule[-0.500pt]{6.263pt}{1.000pt}}
\put(524,626){\rule[-0.500pt]{3.132pt}{1.000pt}}
\put(537,627){\rule[-0.500pt]{6.504pt}{1.000pt}}
\put(564,628){\rule[-0.500pt]{6.263pt}{1.000pt}}
\put(590,629){\rule[-0.500pt]{6.263pt}{1.000pt}}
\put(616,630){\rule[-0.500pt]{6.263pt}{1.000pt}}
\put(642,631){\rule[-0.500pt]{6.263pt}{1.000pt}}
\put(668,632){\rule[-0.500pt]{6.263pt}{1.000pt}}
\put(694,633){\rule[-0.500pt]{6.263pt}{1.000pt}}
\put(720,634){\rule[-0.500pt]{6.263pt}{1.000pt}}
\put(746,635){\rule[-0.500pt]{6.263pt}{1.000pt}}
\put(772,636){\rule[-0.500pt]{6.263pt}{1.000pt}}
\put(798,637){\rule[-0.500pt]{9.395pt}{1.000pt}}
\put(837,638){\rule[-0.500pt]{6.263pt}{1.000pt}}
\put(863,639){\rule[-0.500pt]{6.263pt}{1.000pt}}
\put(889,640){\rule[-0.500pt]{6.263pt}{1.000pt}}
\put(915,641){\rule[-0.500pt]{9.395pt}{1.000pt}}
\put(954,642){\rule[-0.500pt]{6.263pt}{1.000pt}}
\put(980,643){\rule[-0.500pt]{6.263pt}{1.000pt}}
\put(1006,644){\rule[-0.500pt]{9.395pt}{1.000pt}}
\put(1045,645){\rule[-0.500pt]{6.263pt}{1.000pt}}
\put(1071,646){\rule[-0.500pt]{6.263pt}{1.000pt}}
\put(1097,647){\rule[-0.500pt]{6.263pt}{1.000pt}}
\put(1123,648){\rule[-0.500pt]{9.636pt}{1.000pt}}
\put(1163,649){\rule[-0.500pt]{6.263pt}{1.000pt}}
\put(1189,650){\rule[-0.500pt]{6.263pt}{1.000pt}}
\put(1215,651){\rule[-0.500pt]{6.263pt}{1.000pt}}
\put(1241,652){\rule[-0.500pt]{6.263pt}{1.000pt}}
\put(1267,653){\rule[-0.500pt]{6.263pt}{1.000pt}}
\put(1293,654){\rule[-0.500pt]{3.132pt}{1.000pt}}
\put(1306,655){\rule[-0.500pt]{6.263pt}{1.000pt}}
\put(1332,656){\rule[-0.500pt]{6.263pt}{1.000pt}}
\put(1358,657){\rule[-0.500pt]{3.132pt}{1.000pt}}
\put(1371,658){\rule[-0.500pt]{3.132pt}{1.000pt}}
\put(1384,659){\rule[-0.500pt]{3.132pt}{1.000pt}}
\put(1397,660){\rule[-0.500pt]{3.132pt}{1.000pt}}
\put(1410,661){\rule[-0.500pt]{3.132pt}{1.000pt}}
\put(1423,662){\rule[-0.500pt]{1.044pt}{1.000pt}}
\put(1427,663){\rule[-0.500pt]{1.044pt}{1.000pt}}
\put(1431,664){\rule[-0.500pt]{1.044pt}{1.000pt}}
\put(264,747){\usebox{\plotpoint}}
\put(264,747){\rule[-0.500pt]{3.132pt}{1.000pt}}
\put(277,746){\rule[-0.500pt]{1.566pt}{1.000pt}}
\put(283,745){\rule[-0.500pt]{1.566pt}{1.000pt}}
\put(290,744){\rule[-0.500pt]{3.132pt}{1.000pt}}
\put(303,743){\rule[-0.500pt]{3.132pt}{1.000pt}}
\put(316,742){\rule[-0.500pt]{1.566pt}{1.000pt}}
\put(322,741){\rule[-0.500pt]{1.566pt}{1.000pt}}
\put(329,740){\rule[-0.500pt]{3.132pt}{1.000pt}}
\put(342,739){\rule[-0.500pt]{3.132pt}{1.000pt}}
\put(355,738){\rule[-0.500pt]{3.132pt}{1.000pt}}
\put(368,737){\rule[-0.500pt]{3.132pt}{1.000pt}}
\put(381,736){\rule[-0.500pt]{3.132pt}{1.000pt}}
\put(394,735){\rule[-0.500pt]{3.132pt}{1.000pt}}
\put(407,734){\rule[-0.500pt]{3.132pt}{1.000pt}}
\put(420,733){\rule[-0.500pt]{3.132pt}{1.000pt}}
\put(433,732){\rule[-0.500pt]{3.132pt}{1.000pt}}
\put(446,731){\rule[-0.500pt]{3.132pt}{1.000pt}}
\put(459,730){\rule[-0.500pt]{6.263pt}{1.000pt}}
\put(485,729){\rule[-0.500pt]{3.132pt}{1.000pt}}
\put(498,728){\rule[-0.500pt]{3.132pt}{1.000pt}}
\put(511,727){\rule[-0.500pt]{3.132pt}{1.000pt}}
\put(524,726){\rule[-0.500pt]{6.263pt}{1.000pt}}
\put(550,725){\rule[-0.500pt]{3.373pt}{1.000pt}}
\put(564,724){\rule[-0.500pt]{3.132pt}{1.000pt}}
\put(577,723){\rule[-0.500pt]{6.263pt}{1.000pt}}
\put(603,722){\rule[-0.500pt]{3.132pt}{1.000pt}}
\put(616,721){\rule[-0.500pt]{3.132pt}{1.000pt}}
\put(629,720){\rule[-0.500pt]{6.263pt}{1.000pt}}
\put(655,719){\rule[-0.500pt]{3.132pt}{1.000pt}}
\put(668,718){\rule[-0.500pt]{6.263pt}{1.000pt}}
\put(694,717){\rule[-0.500pt]{3.132pt}{1.000pt}}
\put(707,716){\rule[-0.500pt]{6.263pt}{1.000pt}}
\put(733,715){\rule[-0.500pt]{6.263pt}{1.000pt}}
\put(759,714){\rule[-0.500pt]{3.132pt}{1.000pt}}
\put(772,713){\rule[-0.500pt]{6.263pt}{1.000pt}}
\put(798,712){\rule[-0.500pt]{3.132pt}{1.000pt}}
\put(811,711){\rule[-0.500pt]{6.263pt}{1.000pt}}
\put(837,710){\rule[-0.500pt]{3.132pt}{1.000pt}}
\put(850,709){\rule[-0.500pt]{6.263pt}{1.000pt}}
\put(876,708){\rule[-0.500pt]{6.263pt}{1.000pt}}
\put(902,707){\rule[-0.500pt]{3.132pt}{1.000pt}}
\put(915,706){\rule[-0.500pt]{6.263pt}{1.000pt}}
\put(941,705){\rule[-0.500pt]{6.263pt}{1.000pt}}
\put(967,704){\rule[-0.500pt]{3.132pt}{1.000pt}}
\put(980,703){\rule[-0.500pt]{6.263pt}{1.000pt}}
\put(1006,702){\rule[-0.500pt]{3.132pt}{1.000pt}}
\put(1019,701){\rule[-0.500pt]{6.263pt}{1.000pt}}
\put(1045,700){\rule[-0.500pt]{3.132pt}{1.000pt}}
\put(1058,699){\rule[-0.500pt]{6.263pt}{1.000pt}}
\put(1084,698){\rule[-0.500pt]{3.132pt}{1.000pt}}
\put(1097,697){\rule[-0.500pt]{6.263pt}{1.000pt}}
\put(1123,696){\rule[-0.500pt]{3.132pt}{1.000pt}}
\put(1136,695){\rule[-0.500pt]{6.504pt}{1.000pt}}
\put(1163,694){\rule[-0.500pt]{3.132pt}{1.000pt}}
\put(1176,693){\rule[-0.500pt]{3.132pt}{1.000pt}}
\put(1189,692){\rule[-0.500pt]{6.263pt}{1.000pt}}
\put(1215,691){\rule[-0.500pt]{3.132pt}{1.000pt}}
\put(1228,690){\rule[-0.500pt]{3.132pt}{1.000pt}}
\put(1241,689){\rule[-0.500pt]{6.263pt}{1.000pt}}
\put(1267,688){\rule[-0.500pt]{3.132pt}{1.000pt}}
\put(1280,687){\rule[-0.500pt]{3.132pt}{1.000pt}}
\put(1293,686){\rule[-0.500pt]{3.132pt}{1.000pt}}
\put(1306,685){\rule[-0.500pt]{3.132pt}{1.000pt}}
\put(1319,684){\rule[-0.500pt]{3.132pt}{1.000pt}}
\put(1332,683){\rule[-0.500pt]{3.132pt}{1.000pt}}
\put(1345,682){\rule[-0.500pt]{1.566pt}{1.000pt}}
\put(1351,681){\rule[-0.500pt]{1.566pt}{1.000pt}}
\put(1358,680){\rule[-0.500pt]{3.132pt}{1.000pt}}
\put(1371,679){\rule[-0.500pt]{3.132pt}{1.000pt}}
\put(1384,678){\rule[-0.500pt]{1.566pt}{1.000pt}}
\put(1390,677){\rule[-0.500pt]{1.566pt}{1.000pt}}
\put(1397,676){\rule[-0.500pt]{1.566pt}{1.000pt}}
\put(1403,675){\rule[-0.500pt]{1.566pt}{1.000pt}}
\put(1410,674){\rule[-0.500pt]{1.566pt}{1.000pt}}
\put(1416,673){\rule[-0.500pt]{1.566pt}{1.000pt}}
\put(1423,672){\usebox{\plotpoint}}
\put(1424,671){\usebox{\plotpoint}}
\put(1426,670){\usebox{\plotpoint}}
\put(1428,669){\usebox{\plotpoint}}
\put(1430,668){\usebox{\plotpoint}}
\put(1432,667){\usebox{\plotpoint}}
\put(1434,666){\usebox{\plotpoint}}
\put(264,695){\usebox{\plotpoint}}
\put(264,695){\rule[-0.500pt]{12.527pt}{1.000pt}}
\put(316,694){\rule[-0.500pt]{15.658pt}{1.000pt}}
\put(381,693){\rule[-0.500pt]{15.658pt}{1.000pt}}
\put(446,692){\rule[-0.500pt]{15.658pt}{1.000pt}}
\put(511,691){\rule[-0.500pt]{15.899pt}{1.000pt}}
\put(577,690){\rule[-0.500pt]{15.658pt}{1.000pt}}
\put(642,689){\rule[-0.500pt]{15.658pt}{1.000pt}}
\put(707,688){\rule[-0.500pt]{12.527pt}{1.000pt}}
\put(759,687){\rule[-0.500pt]{15.658pt}{1.000pt}}
\put(824,686){\rule[-0.500pt]{12.527pt}{1.000pt}}
\put(876,685){\rule[-0.500pt]{12.527pt}{1.000pt}}
\put(928,684){\rule[-0.500pt]{12.527pt}{1.000pt}}
\put(980,683){\rule[-0.500pt]{12.527pt}{1.000pt}}
\put(1032,682){\rule[-0.500pt]{9.395pt}{1.000pt}}
\put(1071,681){\rule[-0.500pt]{12.527pt}{1.000pt}}
\put(1123,680){\rule[-0.500pt]{9.636pt}{1.000pt}}
\put(1163,679){\rule[-0.500pt]{9.395pt}{1.000pt}}
\put(1202,678){\rule[-0.500pt]{9.395pt}{1.000pt}}
\put(1241,677){\rule[-0.500pt]{6.263pt}{1.000pt}}
\put(1267,676){\rule[-0.500pt]{9.395pt}{1.000pt}}
\put(1306,675){\rule[-0.500pt]{6.263pt}{1.000pt}}
\put(1332,674){\rule[-0.500pt]{6.263pt}{1.000pt}}
\put(1358,673){\rule[-0.500pt]{3.132pt}{1.000pt}}
\put(1371,672){\rule[-0.500pt]{6.263pt}{1.000pt}}
\put(1397,671){\rule[-0.500pt]{3.132pt}{1.000pt}}
\put(1410,670){\rule[-0.500pt]{3.132pt}{1.000pt}}
\put(1423,669){\usebox{\plotpoint}}
\put(1426,668){\usebox{\plotpoint}}
\put(1429,667){\usebox{\plotpoint}}
\put(1432,666){\usebox{\plotpoint}}
\put(264,172){\usebox{\plotpoint}}
\put(264,172){\usebox{\plotpoint}}
\put(265,175){\usebox{\plotpoint}}
\put(266,179){\usebox{\plotpoint}}
\put(267,183){\usebox{\plotpoint}}
\put(268,186){\usebox{\plotpoint}}
\put(269,190){\usebox{\plotpoint}}
\put(270,194){\usebox{\plotpoint}}
\put(271,197){\usebox{\plotpoint}}
\put(272,201){\usebox{\plotpoint}}
\put(273,205){\usebox{\plotpoint}}
\put(274,208){\usebox{\plotpoint}}
\put(275,212){\usebox{\plotpoint}}
\put(276,216){\usebox{\plotpoint}}
\put(277,219){\usebox{\plotpoint}}
\put(278,223){\usebox{\plotpoint}}
\put(279,226){\usebox{\plotpoint}}
\put(280,229){\usebox{\plotpoint}}
\put(281,232){\usebox{\plotpoint}}
\put(282,235){\usebox{\plotpoint}}
\put(283,238){\usebox{\plotpoint}}
\put(284,241){\usebox{\plotpoint}}
\put(285,244){\usebox{\plotpoint}}
\put(286,247){\usebox{\plotpoint}}
\put(287,250){\usebox{\plotpoint}}
\put(288,253){\usebox{\plotpoint}}
\put(289,256){\usebox{\plotpoint}}
\put(290,259){\usebox{\plotpoint}}
\put(291,261){\usebox{\plotpoint}}
\put(292,264){\usebox{\plotpoint}}
\put(293,266){\usebox{\plotpoint}}
\put(294,269){\usebox{\plotpoint}}
\put(295,272){\usebox{\plotpoint}}
\put(296,274){\usebox{\plotpoint}}
\put(297,277){\usebox{\plotpoint}}
\put(298,279){\usebox{\plotpoint}}
\put(299,282){\usebox{\plotpoint}}
\put(300,285){\usebox{\plotpoint}}
\put(301,287){\usebox{\plotpoint}}
\put(302,290){\usebox{\plotpoint}}
\put(303,293){\usebox{\plotpoint}}
\put(304,295){\usebox{\plotpoint}}
\put(305,297){\usebox{\plotpoint}}
\put(306,299){\usebox{\plotpoint}}
\put(307,301){\usebox{\plotpoint}}
\put(308,304){\usebox{\plotpoint}}
\put(309,306){\usebox{\plotpoint}}
\put(310,308){\usebox{\plotpoint}}
\put(311,310){\usebox{\plotpoint}}
\put(312,313){\usebox{\plotpoint}}
\put(313,315){\usebox{\plotpoint}}
\put(314,317){\usebox{\plotpoint}}
\put(315,319){\usebox{\plotpoint}}
\put(316,322){\usebox{\plotpoint}}
\put(317,323){\usebox{\plotpoint}}
\put(318,325){\usebox{\plotpoint}}
\put(319,327){\usebox{\plotpoint}}
\put(320,329){\usebox{\plotpoint}}
\put(321,331){\usebox{\plotpoint}}
\put(322,333){\usebox{\plotpoint}}
\put(323,335){\usebox{\plotpoint}}
\put(324,337){\usebox{\plotpoint}}
\put(325,339){\usebox{\plotpoint}}
\put(326,341){\usebox{\plotpoint}}
\put(327,343){\usebox{\plotpoint}}
\put(328,345){\usebox{\plotpoint}}
\put(329,346){\usebox{\plotpoint}}
\put(330,348){\usebox{\plotpoint}}
\put(331,350){\usebox{\plotpoint}}
\put(332,351){\usebox{\plotpoint}}
\put(333,353){\usebox{\plotpoint}}
\put(334,355){\usebox{\plotpoint}}
\put(335,356){\usebox{\plotpoint}}
\put(336,358){\usebox{\plotpoint}}
\put(337,359){\usebox{\plotpoint}}
\put(338,361){\usebox{\plotpoint}}
\put(339,363){\usebox{\plotpoint}}
\put(340,364){\usebox{\plotpoint}}
\put(341,366){\usebox{\plotpoint}}
\put(342,368){\usebox{\plotpoint}}
\put(343,369){\usebox{\plotpoint}}
\put(344,370){\usebox{\plotpoint}}
\put(345,372){\usebox{\plotpoint}}
\put(346,373){\usebox{\plotpoint}}
\put(347,375){\usebox{\plotpoint}}
\put(348,376){\usebox{\plotpoint}}
\put(349,378){\usebox{\plotpoint}}
\put(350,379){\usebox{\plotpoint}}
\put(351,381){\usebox{\plotpoint}}
\put(352,382){\usebox{\plotpoint}}
\put(353,384){\usebox{\plotpoint}}
\put(354,385){\usebox{\plotpoint}}
\put(355,387){\usebox{\plotpoint}}
\put(356,388){\usebox{\plotpoint}}
\put(357,389){\usebox{\plotpoint}}
\put(358,390){\usebox{\plotpoint}}
\put(359,392){\usebox{\plotpoint}}
\put(360,393){\usebox{\plotpoint}}
\put(361,394){\usebox{\plotpoint}}
\put(362,396){\usebox{\plotpoint}}
\put(363,397){\usebox{\plotpoint}}
\put(364,398){\usebox{\plotpoint}}
\put(365,400){\usebox{\plotpoint}}
\put(366,401){\usebox{\plotpoint}}
\put(367,402){\usebox{\plotpoint}}
\put(368,403){\usebox{\plotpoint}}
\put(369,405){\usebox{\plotpoint}}
\put(370,406){\usebox{\plotpoint}}
\put(371,407){\usebox{\plotpoint}}
\put(372,408){\usebox{\plotpoint}}
\put(373,410){\usebox{\plotpoint}}
\put(374,411){\usebox{\plotpoint}}
\put(375,412){\usebox{\plotpoint}}
\put(376,413){\usebox{\plotpoint}}
\put(377,415){\usebox{\plotpoint}}
\put(378,416){\usebox{\plotpoint}}
\put(379,417){\usebox{\plotpoint}}
\put(380,418){\usebox{\plotpoint}}
\put(381,420){\usebox{\plotpoint}}
\put(382,421){\usebox{\plotpoint}}
\put(383,422){\usebox{\plotpoint}}
\put(384,423){\usebox{\plotpoint}}
\put(385,424){\usebox{\plotpoint}}
\put(386,425){\usebox{\plotpoint}}
\put(387,426){\usebox{\plotpoint}}
\put(388,427){\usebox{\plotpoint}}
\put(389,428){\usebox{\plotpoint}}
\put(390,429){\usebox{\plotpoint}}
\put(391,430){\usebox{\plotpoint}}
\put(392,431){\usebox{\plotpoint}}
\put(393,432){\usebox{\plotpoint}}
\put(394,433){\usebox{\plotpoint}}
\put(395,434){\usebox{\plotpoint}}
\put(396,435){\usebox{\plotpoint}}
\put(397,436){\usebox{\plotpoint}}
\put(398,437){\usebox{\plotpoint}}
\put(399,438){\usebox{\plotpoint}}
\put(400,439){\usebox{\plotpoint}}
\put(401,440){\usebox{\plotpoint}}
\put(402,441){\usebox{\plotpoint}}
\put(403,442){\usebox{\plotpoint}}
\put(404,443){\usebox{\plotpoint}}
\put(405,444){\usebox{\plotpoint}}
\put(406,445){\usebox{\plotpoint}}
\put(407,446){\usebox{\plotpoint}}
\put(408,447){\usebox{\plotpoint}}
\put(409,448){\usebox{\plotpoint}}
\put(410,449){\usebox{\plotpoint}}
\put(411,450){\usebox{\plotpoint}}
\put(412,451){\usebox{\plotpoint}}
\put(414,452){\usebox{\plotpoint}}
\put(415,453){\usebox{\plotpoint}}
\put(416,454){\usebox{\plotpoint}}
\put(417,455){\usebox{\plotpoint}}
\put(418,456){\usebox{\plotpoint}}
\put(420,457){\usebox{\plotpoint}}
\put(421,458){\usebox{\plotpoint}}
\put(422,459){\usebox{\plotpoint}}
\put(423,460){\usebox{\plotpoint}}
\put(425,461){\usebox{\plotpoint}}
\put(426,462){\usebox{\plotpoint}}
\put(427,463){\usebox{\plotpoint}}
\put(429,464){\usebox{\plotpoint}}
\put(430,465){\usebox{\plotpoint}}
\put(431,466){\usebox{\plotpoint}}
\put(432,467){\usebox{\plotpoint}}
\put(434,468){\usebox{\plotpoint}}
\put(435,469){\usebox{\plotpoint}}
\put(436,470){\usebox{\plotpoint}}
\put(438,471){\usebox{\plotpoint}}
\put(439,472){\usebox{\plotpoint}}
\put(440,473){\usebox{\plotpoint}}
\put(442,474){\usebox{\plotpoint}}
\put(443,475){\usebox{\plotpoint}}
\put(444,476){\usebox{\plotpoint}}
\put(445,477){\usebox{\plotpoint}}
\put(447,478){\usebox{\plotpoint}}
\put(449,479){\usebox{\plotpoint}}
\put(450,480){\usebox{\plotpoint}}
\put(452,481){\usebox{\plotpoint}}
\put(454,482){\usebox{\plotpoint}}
\put(455,483){\usebox{\plotpoint}}
\put(457,484){\usebox{\plotpoint}}
\put(459,485){\usebox{\plotpoint}}
\put(460,486){\usebox{\plotpoint}}
\put(462,487){\usebox{\plotpoint}}
\put(463,488){\usebox{\plotpoint}}
\put(465,489){\usebox{\plotpoint}}
\put(467,490){\usebox{\plotpoint}}
\put(468,491){\usebox{\plotpoint}}
\put(470,492){\usebox{\plotpoint}}
\put(472,493){\usebox{\plotpoint}}
\put(473,494){\usebox{\plotpoint}}
\put(475,495){\usebox{\plotpoint}}
\put(476,496){\usebox{\plotpoint}}
\put(478,497){\usebox{\plotpoint}}
\put(480,498){\usebox{\plotpoint}}
\put(481,499){\usebox{\plotpoint}}
\put(483,500){\usebox{\plotpoint}}
\put(485,501){\usebox{\plotpoint}}
\put(487,502){\usebox{\plotpoint}}
\put(489,503){\usebox{\plotpoint}}
\put(491,504){\usebox{\plotpoint}}
\put(493,505){\usebox{\plotpoint}}
\put(495,506){\usebox{\plotpoint}}
\put(497,507){\usebox{\plotpoint}}
\put(499,508){\usebox{\plotpoint}}
\put(501,509){\usebox{\plotpoint}}
\put(503,510){\usebox{\plotpoint}}
\put(505,511){\usebox{\plotpoint}}
\put(507,512){\usebox{\plotpoint}}
\put(509,513){\usebox{\plotpoint}}
\put(511,514){\usebox{\plotpoint}}
\put(513,515){\usebox{\plotpoint}}
\put(515,516){\usebox{\plotpoint}}
\put(517,517){\usebox{\plotpoint}}
\put(519,518){\usebox{\plotpoint}}
\put(521,519){\usebox{\plotpoint}}
\put(524,520){\usebox{\plotpoint}}
\put(526,521){\usebox{\plotpoint}}
\put(529,522){\usebox{\plotpoint}}
\put(531,523){\usebox{\plotpoint}}
\put(534,524){\usebox{\plotpoint}}
\put(536,525){\usebox{\plotpoint}}
\put(539,526){\usebox{\plotpoint}}
\put(542,527){\usebox{\plotpoint}}
\put(544,528){\usebox{\plotpoint}}
\put(547,529){\usebox{\plotpoint}}
\put(549,530){\usebox{\plotpoint}}
\put(552,531){\usebox{\plotpoint}}
\put(555,532){\usebox{\plotpoint}}
\put(558,533){\usebox{\plotpoint}}
\put(561,534){\usebox{\plotpoint}}
\put(563,535){\usebox{\plotpoint}}
\put(566,536){\usebox{\plotpoint}}
\put(569,537){\usebox{\plotpoint}}
\put(571,538){\usebox{\plotpoint}}
\put(574,539){\usebox{\plotpoint}}
\put(576,540){\usebox{\plotpoint}}
\put(580,541){\usebox{\plotpoint}}
\put(583,542){\usebox{\plotpoint}}
\put(586,543){\usebox{\plotpoint}}
\put(590,544){\usebox{\plotpoint}}
\put(593,545){\usebox{\plotpoint}}
\put(596,546){\usebox{\plotpoint}}
\put(599,547){\usebox{\plotpoint}}
\put(603,548){\usebox{\plotpoint}}
\put(606,549){\usebox{\plotpoint}}
\put(609,550){\usebox{\plotpoint}}
\put(612,551){\usebox{\plotpoint}}
\put(616,552){\usebox{\plotpoint}}
\put(619,553){\usebox{\plotpoint}}
\put(622,554){\usebox{\plotpoint}}
\put(625,555){\usebox{\plotpoint}}
\put(629,556){\rule[-0.500pt]{1.044pt}{1.000pt}}
\put(633,557){\rule[-0.500pt]{1.044pt}{1.000pt}}
\put(637,558){\rule[-0.500pt]{1.044pt}{1.000pt}}
\put(641,559){\rule[-0.500pt]{1.044pt}{1.000pt}}
\put(646,560){\rule[-0.500pt]{1.044pt}{1.000pt}}
\put(650,561){\rule[-0.500pt]{1.044pt}{1.000pt}}
\put(654,562){\rule[-0.500pt]{1.044pt}{1.000pt}}
\put(659,563){\rule[-0.500pt]{1.044pt}{1.000pt}}
\put(663,564){\rule[-0.500pt]{1.044pt}{1.000pt}}
\put(667,565){\rule[-0.500pt]{1.044pt}{1.000pt}}
\put(672,566){\rule[-0.500pt]{1.044pt}{1.000pt}}
\put(676,567){\rule[-0.500pt]{1.044pt}{1.000pt}}
\put(680,568){\rule[-0.500pt]{1.044pt}{1.000pt}}
\put(685,569){\rule[-0.500pt]{1.044pt}{1.000pt}}
\put(689,570){\rule[-0.500pt]{1.044pt}{1.000pt}}
\put(693,571){\rule[-0.500pt]{1.044pt}{1.000pt}}
\put(698,572){\rule[-0.500pt]{1.044pt}{1.000pt}}
\put(702,573){\rule[-0.500pt]{1.044pt}{1.000pt}}
\put(706,574){\rule[-0.500pt]{1.566pt}{1.000pt}}
\put(713,575){\rule[-0.500pt]{1.566pt}{1.000pt}}
\put(720,576){\rule[-0.500pt]{1.044pt}{1.000pt}}
\put(724,577){\rule[-0.500pt]{1.044pt}{1.000pt}}
\put(728,578){\rule[-0.500pt]{1.044pt}{1.000pt}}
\put(732,579){\rule[-0.500pt]{1.566pt}{1.000pt}}
\put(739,580){\rule[-0.500pt]{1.566pt}{1.000pt}}
\put(746,581){\rule[-0.500pt]{1.566pt}{1.000pt}}
\put(752,582){\rule[-0.500pt]{1.566pt}{1.000pt}}
\put(759,583){\rule[-0.500pt]{1.044pt}{1.000pt}}
\put(763,584){\rule[-0.500pt]{1.044pt}{1.000pt}}
\put(767,585){\rule[-0.500pt]{1.044pt}{1.000pt}}
\put(771,586){\rule[-0.500pt]{1.566pt}{1.000pt}}
\put(778,587){\rule[-0.500pt]{1.566pt}{1.000pt}}
\put(785,588){\rule[-0.500pt]{1.566pt}{1.000pt}}
\put(791,589){\rule[-0.500pt]{1.566pt}{1.000pt}}
\put(798,590){\rule[-0.500pt]{1.566pt}{1.000pt}}
\put(804,591){\rule[-0.500pt]{1.566pt}{1.000pt}}
\put(811,592){\rule[-0.500pt]{1.566pt}{1.000pt}}
\put(817,593){\rule[-0.500pt]{1.566pt}{1.000pt}}
\put(824,594){\rule[-0.500pt]{3.132pt}{1.000pt}}
\put(837,595){\rule[-0.500pt]{1.566pt}{1.000pt}}
\put(843,596){\rule[-0.500pt]{1.566pt}{1.000pt}}
\put(850,597){\rule[-0.500pt]{1.566pt}{1.000pt}}
\put(856,598){\rule[-0.500pt]{1.566pt}{1.000pt}}
\put(863,599){\rule[-0.500pt]{3.132pt}{1.000pt}}
\put(876,600){\rule[-0.500pt]{1.566pt}{1.000pt}}
\put(882,601){\rule[-0.500pt]{1.566pt}{1.000pt}}
\put(889,602){\rule[-0.500pt]{1.566pt}{1.000pt}}
\put(895,603){\rule[-0.500pt]{1.566pt}{1.000pt}}
\put(902,604){\rule[-0.500pt]{3.132pt}{1.000pt}}
\put(915,605){\rule[-0.500pt]{1.566pt}{1.000pt}}
\put(921,606){\rule[-0.500pt]{1.566pt}{1.000pt}}
\put(928,607){\rule[-0.500pt]{3.132pt}{1.000pt}}
\put(941,608){\rule[-0.500pt]{1.566pt}{1.000pt}}
\put(947,609){\rule[-0.500pt]{1.566pt}{1.000pt}}
\put(954,610){\rule[-0.500pt]{3.132pt}{1.000pt}}
\put(967,611){\rule[-0.500pt]{3.132pt}{1.000pt}}
\put(980,612){\rule[-0.500pt]{1.566pt}{1.000pt}}
\put(986,613){\rule[-0.500pt]{1.566pt}{1.000pt}}
\put(993,614){\rule[-0.500pt]{3.132pt}{1.000pt}}
\put(1006,615){\rule[-0.500pt]{3.132pt}{1.000pt}}
\put(1019,616){\rule[-0.500pt]{1.566pt}{1.000pt}}
\put(1025,617){\rule[-0.500pt]{1.566pt}{1.000pt}}
\put(1032,618){\rule[-0.500pt]{3.132pt}{1.000pt}}
\put(1045,619){\rule[-0.500pt]{3.132pt}{1.000pt}}
\put(1058,620){\rule[-0.500pt]{3.132pt}{1.000pt}}
\put(1071,621){\rule[-0.500pt]{3.132pt}{1.000pt}}
\put(1084,622){\rule[-0.500pt]{1.566pt}{1.000pt}}
\put(1090,623){\rule[-0.500pt]{1.566pt}{1.000pt}}
\put(1097,624){\rule[-0.500pt]{3.132pt}{1.000pt}}
\put(1110,625){\rule[-0.500pt]{3.132pt}{1.000pt}}
\put(1123,626){\rule[-0.500pt]{3.132pt}{1.000pt}}
\put(1136,627){\rule[-0.500pt]{3.373pt}{1.000pt}}
\put(1150,628){\rule[-0.500pt]{1.566pt}{1.000pt}}
\put(1156,629){\rule[-0.500pt]{1.566pt}{1.000pt}}
\put(1163,630){\rule[-0.500pt]{3.132pt}{1.000pt}}
\put(1176,631){\rule[-0.500pt]{3.132pt}{1.000pt}}
\put(1189,632){\rule[-0.500pt]{3.132pt}{1.000pt}}
\put(1202,633){\rule[-0.500pt]{3.132pt}{1.000pt}}
\put(1215,634){\rule[-0.500pt]{3.132pt}{1.000pt}}
\put(1228,635){\rule[-0.500pt]{1.566pt}{1.000pt}}
\put(1234,636){\rule[-0.500pt]{1.566pt}{1.000pt}}
\put(1241,637){\rule[-0.500pt]{3.132pt}{1.000pt}}
\put(1254,638){\rule[-0.500pt]{3.132pt}{1.000pt}}
\put(1267,639){\rule[-0.500pt]{3.132pt}{1.000pt}}
\put(1280,640){\rule[-0.500pt]{1.566pt}{1.000pt}}
\put(1286,641){\rule[-0.500pt]{1.566pt}{1.000pt}}
\put(1293,642){\rule[-0.500pt]{3.132pt}{1.000pt}}
\put(1306,643){\rule[-0.500pt]{3.132pt}{1.000pt}}
\put(1319,644){\rule[-0.500pt]{1.566pt}{1.000pt}}
\put(1325,645){\rule[-0.500pt]{1.566pt}{1.000pt}}
\put(1332,646){\rule[-0.500pt]{3.132pt}{1.000pt}}
\put(1345,647){\rule[-0.500pt]{1.566pt}{1.000pt}}
\put(1351,648){\rule[-0.500pt]{1.566pt}{1.000pt}}
\put(1358,649){\rule[-0.500pt]{3.132pt}{1.000pt}}
\put(1371,650){\rule[-0.500pt]{1.566pt}{1.000pt}}
\put(1377,651){\rule[-0.500pt]{1.566pt}{1.000pt}}
\put(1384,652){\rule[-0.500pt]{1.566pt}{1.000pt}}
\put(1390,653){\rule[-0.500pt]{1.566pt}{1.000pt}}
\put(1397,654){\rule[-0.500pt]{1.566pt}{1.000pt}}
\put(1403,655){\rule[-0.500pt]{1.566pt}{1.000pt}}
\put(1410,656){\rule[-0.500pt]{1.044pt}{1.000pt}}
\put(1414,657){\rule[-0.500pt]{1.044pt}{1.000pt}}
\put(1418,658){\rule[-0.500pt]{1.044pt}{1.000pt}}
\put(1423,659){\usebox{\plotpoint}}
\put(1425,660){\usebox{\plotpoint}}
\put(1427,661){\usebox{\plotpoint}}
\put(1429,662){\usebox{\plotpoint}}
\put(1431,663){\usebox{\plotpoint}}
\put(1433,664){\usebox{\plotpoint}}
\put(1435,665){\usebox{\plotpoint}}
\end{picture}

%% file: ms.bbl
\begin{thebibliography}{99}
\bibitem{note1}
Experimentally a given strength of pinning potential can be obtained by 
adjusting the difference in the solid-liquid and solid-gas surface energy, 
$A^{'}_p (y,z) = \gamma_{SL} (y,z) - \gamma_{SG} (y,z)$. From Young's
 relation \cite{Young} $\gamma_{SL} (y,z) + \gamma_{LG} \cos{ \theta (y,z)} 
= \gamma_{SG} (y,z)$ the strength of pinning potential is seen to correspond 
to given macroscopic contact angle $\theta (y,z)$.
\end{thebibliography}
